\documentclass[prd, aps, 10pt, twocolumn, nofootinbib, superscriptaddress, preprintnumbers, balancelastpage, longbibliography, floatfix]{revtex4-2}
\usepackage[english]{babel}

\usepackage[dvipsnames]{xcolor}
\usepackage{amsmath, amssymb, mathtools, physics, xfrac, graphicx, afterpage, subfigure, rotating, multirow, tabularx, booktabs, fancyhdr, theorem, moreverb, euscript, psfrag, slashed, mathtools, makecell, adjustbox, dcolumn, bm, graphics, hyperref, lettrine, aas_macros, color, tikz, soul, float}
\usepackage[utf8]{inputenc}
\usetikzlibrary{decorations.pathmorphing, fadings, decorations.text, arrows.meta}

\definecolor{darkblue}{rgb}{0.0,0.0,0.75}
\definecolor{darkred}{rgb}{0.6,0.0,0}
\definecolor{darkgreen}{rgb}{0.0,0.6,0.}
\definecolor{nice}{rgb}{0.8,0, 0.8}

\hypersetup{
     colorlinks = true,
     citecolor  = Green,
     urlcolor   = Green,
     linkcolor  = Green
}

\newcommand{\hp}[0]{H$^+_3\;$}
\newcommand{\fatm}{f_\text{atm}}
\newcommand{\fint}{f_\text{int}}

\input Zallman.fd

\LettrineTextFont{\itshape}
\begin{document}

\preprint{SLAC-PUB-250609}
\preprint{MIT--CTP/5749}

\title{Complementary Planetary Spectroscopy Probes of Dark Matter}

\author{Carlos Blanco}
\thanks{\href{mailto:carlosblanco2718@princeton.edu}{carlosblanco2718@princeton.edu}, \href{https://orcid.org/0000-0001-8971-834X}{0000-0001-8971-834X}}
\affiliation{Institute for Gravitation and the Cosmos, The Pennsylvania State University, University Park, PA 16802, USA}
\affiliation{Princeton University, Department of Physics, Princeton, NJ 08544, USA}
\affiliation{Stockholm University and The Oskar Klein Centre for Cosmoparticle Physics, Alba Nova, 10691 Stockholm, Sweden}

\author{Rebecca~K.~Leane}
\thanks{\href{mailto:rleane@slac.stanford.edu}{rleane@slac.stanford.edu}, \href{https://orcid.org/0000-0002-1287-8780}{0000-0002-1287-8780}}
\affiliation{Particle Theory Group, SLAC National Accelerator Laboratory, Stanford, CA 94305, USA}
\affiliation{Kavli Institute for Particle Astrophysics and Cosmology, Stanford University, Stanford, CA 94305, USA}

\author{Marianne Moore}
\thanks{\href{mailto:mamoore@mit.edu}{mamoore@mit.edu}, \href{https://orcid.org/0000-0002-6424-0594}{0000-0002-6424-0594}}
\affiliation{Center for Theoretical Physics -- a Leinweber Institute, Massachusetts Institute of Technology, Cambridge, MA 02139, USA}
\affiliation{Department of Physics, Harvard University, Cambridge, MA 02138, USA}

\author{Joshua Tong}
\thanks{\href{mailto:joshtong@stanford.edu}{joshtong@stanford.edu}, \href{https://orcid.org/0000-0002-6818-055X}{0000-0002-6818-055X}}
\affiliation{Particle Theory Group, SLAC National Accelerator Laboratory, Stanford, CA 94305, USA}
\affiliation{Kavli Institute for Particle Astrophysics and Cosmology, Stanford University, Stanford, CA 94305, USA}

\date{\today}

\begin{abstract}
We investigate dark matter (DM) interactions via spectroscopic signatures of energy injection in planetary environments. We develop a general framework to account for how DM energy injection signals depend on the DM spatial distribution, planetary structure, and DM energy deposition profile. We combine UV airglow data on the Solar System's gas giants from the Voyager and New Horizons flybys, and ionospheric measurements from AMS-02 and ELFIN CubeSat on Earth, with internal heat flow data from Cassini, Voyager, and terrestrial boreholes, to constrain DM-nucleon scattering across both heavy and light mediator scenarios. We show that Earth, gas giants, and ice giants probe complementary DM masses and mediator properties, and forecast the reach of a free-floating Super-Jupiter. These results establish planetary spectroscopy as a powerful and versatile probe of the dark sector, complementary to direct detection, cosmology, and collider searches.
\end{abstract}

\maketitle

\lettrine{F}{rom swirling storms} to faint nightglow, planets are shaped by the energy that moves through them. Long after the last sunlight fades, planets continue to glow with their own subtle signatures, offering clues about the processes unfolding within their atmospheres and cores. The quest to understand the origins of these signals has driven decades of study, resulting in rich, multi-wavelength datasets on planetary atmospheric spectra and internal dynamics.

While many planetary emissions are well understood through known atmospheric and internal processes, new physics, such as dark matter (DM), can also leave imprints in these datasets. DM can scatter off planetary constituents, lose energy, and become gravitationally captured. 
If the captured DM subsequently annihilates, the resulting signatures fall into two broad classes: (1) direct detection of the annihilation products, from electromagnetic sources~\cite{Batell:2009zp,Schuster:2009au,Schuster:2009fc,Meade:2009mu,
Leane:2021ihh, Acevedo:2023xnu, Feng:2015hja, Feng:2016ijc, Leane:2017vag, Niblaeus:2019gjk, Bose:2021cou, HAWC:2018szf, Leane:2021tjj, Chen:2023fgr, Li:2022wix,Linden:2024uph,Leane:2024bvh,Arina:2017sng, Albert:2018jwh, Albert:2018vcq, Nisa:2019mpb, Cuoco:2019mlb, Serini:2020yhb, Mazziotta:2020foa, Bell:2021pyy, Cermeno:2018qgu, Krall:2017xij, Moore:2024mot, Acevedo:2024ttq, Bhattacharjee:2025iip,Gupta:2025jte} or neutrinos~\cite{Freese:1985qw, Hooper:2008cf, Bell:2011sn, Nguyen:2022zwb, Lin:2022dbl, Bose:2021yhz, Bose:2023yll, Chauhan:2023zuf, Maity:2023rez, Leane:2017vag,Ansarifard:2024fan, French:2022ccb, Super-Kamiokande:2015xms, Abe:2011ts, ANTARES:2016xuh, ANTARES:2016bxz, IceCube:2016yoy, IceCube:2020wxa, Pospelov:2023mlz, McKeen:2023ztq, Allahverdi:2016fvl, Garani:2017jcj, Bhattacharjee:2023qfi, Acevedo:2024ttq, Krishna:2025ncv, Nguyen:2025ygc, Bose:2024wsh, Robles:2024tdh, Berlin:2024lwe,Moore:2024mot},
and (2) indirect signals from the energy deposited into the planetary system, measured through planetary spectroscopy. 
These include the most commonly studied scenario of DM annihilation heating~\cite{Kouvaris:2007ay, deLavallaz:2010wp, Kouvaris:2010vv, Baryakhtar:2017dbj, Bell:2018pkk, Chen:2018ohx, Bell:2023ysh, Hamaguchi:2019oev, Camargo:2019wou, Bell:2019pyc, Garani:2019fpa, Acevedo:2019agu, Joglekar:2019vzy, Joglekar:2020liw, Garani:2020wge, Acuna:2022ouv, Alvarez:2023fjj, Fujiwara:2023hlj, Leane:2020wob, Ilie:2023lbi, John:2023knt, Dasgupta:2019juq, Croon:2023bmu, Kawasaki:1991eu, Abbas:1996kk, Mitra:2004fh, Farrar:2005zd, Mack:2007xj, Adler:2008ky, 2013JPhG...40k5202M, Chauhan:2016joa, Bramante:2019fhi, Bramante:2022pmn, Bell:2021fye, Cappiello:2025yfe, Mack:2007xj, Benito:2024yki, Moskalenko:2006mk, Bertone:2007ae, McCullough:2010ai, Hooper:2010es, Amaro-Seoane:2015uny, Graham:2018efk, Panotopoulos:2020kuo, Ramirez-Quezada:2022uou, Garani:2023esk, Fairbairn:2007bn, Scott:2008ns, Iocco:2008xb, Freese:2008hb, Taoso:2008kw, Sivertsson:2010zm, Freese:2015mta, Ilie:2020iup, Ilie:2020nzp, Lopes:2021jcy, Croon:2023trk, Ilie:2023zfv, Freese:2010re, John:2024thz, 1989ApJ...338...24S, Bramante:2023djs},
as well as more recent ideas including ultraviolet (UV) airglow~\cite{Blanco:2024lqw} and \hp ionization~\cite{Blanco:2023qgi}.

In our companion paper~\cite{Blanco:2024lqw}, we introduced a new signature – UV airglow induced by DM – and demonstrated its potential for testing DM-nucleon scattering. If DM annihilation produces Standard Model (SM) particles in the atmosphere, these can excite molecular hydrogen, leading to de-excitation via UV photon emission. Using nightside UV measurements from space probes such as Voyager and New Horizons, we showed that the Solar System’s giant planets can place unprecedented constraints on the DM-nucleon scattering cross section. We also demonstrated that future UV observations of a nearby Super-Jupiter could significantly enhance this sensitivity. That analysis assumed that all DM annihilation products deposit their energy in the atmosphere.

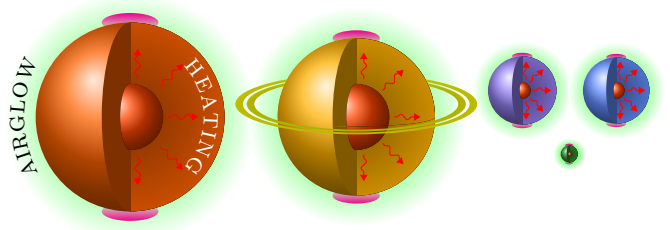
\begin{figure}[t]
    \centering
    \scalebox{.5}{\begin{tikzpicture}[
  snakearrow/.style={
    -Triangle, red, line width=0.8pt,
    decorate, decoration={snake, amplitude=2pt, pre length=2pt, post length=3pt}
  }
]
    \def\rj{2.5} 
    \def\rs{2.084} 
    \def\ru{0.9065} 
    \def\rn{0.88} 
    \def\re{0.2277} 
    \def\dis{1.4} 
    \def\rtiny{0.06}

    \definecolor{colJ}{RGB}{252,97,0}
    \definecolor{colS}{RGB}{255,176,0}
    \definecolor{colU}{RGB}{148,128,242}
    \definecolor{colN}{RGB}{99,143,255}
    \definecolor{colE}{RGB}{51,143,51} 
    \definecolor{colcore}{RGB}{255,69,0}
    \def\ls{\rj+\rs+\dis} 
    \def\lu{\rj+2*\rs+\ru+2*\dis} 
    \def\ln{\rj+2*\rs+2*\ru+\rn+2.5*\dis} 
    \def\le{\rj+2*\rs+1.5*\ru+0.5*\rn+2.25*\dis} 

    \foreach \a/\x/\y in {0.8/0/\rj, 0.8/\ls/\rs} 
    \shade [inner color=green, outer color=white, even odd rule] (\x,0) circle (\y+\a) circle (\y);

    \foreach \a/\x/\y in {0.4/\lu/\ru, 0.4/\ln/\rn} 
    \shade [inner color=green, outer color=white, even odd rule] (\x,{0.5*\dis}) circle (\y+\a) circle (\y);

    \shade [inner color=green, outer color=white, even odd rule] (\le,-{0.7*\dis}) circle (\re+0.2) circle (\re);
    
    \foreach \x/\y in {0/\rj, \ls/\rs} {
    \fill[color=magenta, path fading=north] (\x, -\y) ellipse (\y*0.3 and \y*0.1);
    \fill[color=magenta, path fading=south] (\x, \y) ellipse (\y*0.3 and \y*0.1);
    }

    \foreach \x/\y in {\lu/\ru, \ln/\rn} {
    \fill[color=magenta, path fading=north] (\x, {0.5*\dis-\y}) ellipse (\y*0.3 and \y*0.1);
    \fill[color=magenta, path fading=south] (\x, {0.5*\dis+\y}) ellipse (\y*0.3 and \y*0.1);
    }

    \fill[color=magenta, path fading=north] (\le, {-0.7*\dis-\re}) ellipse (\re*0.5 and \re*0.2);
    \fill[color=magenta, path fading=south] (\le, {-0.7*\dis+\re}) ellipse (\re*0.5 and \re*0.2);

    \foreach \a/\x/\y in {colJ/0/\rj, colS/\ls/\rs} 
    \shade[ball color=\a] (\x,0) circle (\y);
    \foreach \a/\x/\y in {colU/\lu/\ru, colN/\ln/\rn} 
    \shade[ball color=\a] (\x,{0.5*\dis}) circle (\y);
    \shade[ball color=colE] (\le,{-0.7*\dis}) circle (\re);

    \foreach \rad/\dx/\dy/\col in {\rj/0/0/colJ, \ru/\lu/0.5/colU, \rn/\ln/0.5/colN, \re/\le/-0.7/colE, \rs/\ls/0/colS}{
    \begin{scope}[shift={(\dx,{\dy*\dis})}]
        \clip (0,-\rad) rectangle (\rad,\rad);
        \fill[outer color=\col!80!black,inner color=\col!50!black] (0,0) circle (\rad);
    \end{scope}
    \begin{scope}[shift={(\dx,{\dy*\dis})}]
        \clip (0,-\rad) rectangle (-\rad,\rad);
        \fill[\col!50!black] (0,0) ellipse (0.3*\rad cm and \rad cm);
    \end{scope}
    }

    \foreach \rad/\dx/\dy in {\rn/0/0, \re/\lu/0.5, \re/\ln/0.5, \rtiny/\le/-0.7,\rn/\ls/0}{
    \begin{scope}[shift={(\dx,{\dy*\dis})}]
        \clip (90:\rad) arc (90:270:0.3*\rad cm and \rad cm) arc (-90:90:\rad);
        \fill[ball color=colcore,shading=ball] (0,0) circle (\rad cm);
    \end{scope}    
    }

    \draw[snakearrow] (1,0) -- ++(0.8,0);
    \draw[snakearrow] (0.2,1) -- ++(0,0.8);
    \draw[snakearrow] (0.8,0.8) -- ++(0.6,0.6);
    \draw[snakearrow] (0.2,-1) -- ++(0,-0.8);
    \draw[snakearrow] (0.8,-0.8) -- ++(0.6,-0.6);

    \foreach \dx in {\lu, \ln}{
    \begin{scope}[shift={(\dx,{0.5*\dis})}, scale=0.7, transform shape]
        \draw[snakearrow] (0.4,0) -- ++(0.7,0);
        \draw[snakearrow] (0.15,0.4) -- ++(0,0.7);
        \draw[snakearrow] (0.3,0.3) -- ++(0.5,0.5);
        \draw[snakearrow] (0.15,-0.4) -- ++(0,-0.7);
        \draw[snakearrow] (0.3,-0.3) -- ++(0.5,-0.5);
    \end{scope}
    }
    
    \draw [decorate, decoration={raise=0.05cm, text along path, text align=center, text={|\sc\huge|airglow}}] ({-0.8*\rj},{-0.8*\rj}) to [bend left=40]  ({-0.8*\rj},{0.8*\rj});
    \draw[decorate, decoration={raise=0.1cm, text along path, text align=center, text={|\sc\huge\color{white}|heating}}] (45:{0.74*\rj}) arc (45:-45:{0.74*\rj});

    \def\ringColorInner{black!30!olive}
    \def\ringColorOuter{yellow!50!olive}
    \def\ringX{\ls}
    \def\ringY{0.34}
    \def\innerRadius{1.9}
    \def\ringWidth{0.8}

    \path[inner color=\ringColorInner, outer color=\ringColorOuter, even odd rule]    
        (\ringX, \ringY) circle [x radius = \innerRadius + \ringWidth, y radius = (\innerRadius + \ringWidth)/4]
        (\ringX, \ringY) circle [x radius = \innerRadius + \ringWidth + 0.2, y radius = (\innerRadius + \ringWidth + 0.2)/4]
        (\ringX, \ringY) circle [x radius = \innerRadius + \ringWidth + 0.3, y radius = (\innerRadius + \ringWidth + 0.3)/4]
        (\ringX, \ringY) circle [x radius = \innerRadius + \ringWidth + 0.5, y radius = (\innerRadius + \ringWidth + 0.5)/4];

    \begin{scope}[shift={(\ls,0)}]
    \clip (-\rs,0) arc[start angle=180, end angle=0, x radius=\rs, y radius=-0.25]
        arc[start angle=0, end angle=180, x radius=\rs, y radius=\rs];
    \shade[ball color=colS] (0,0) circle[radius=\rs];
    \clip (0,-\rs) rectangle (\rs,\rs);
    \fill[line width=0pt, outer color=colS!80!black,inner color=colS!50!black] (0,0) circle (\rs);
    \end{scope}

    \begin{scope}[shift={(\ls,0)}]
        \clip (-\rs,0) arc[start angle=180, end angle=0, x radius=\rs, y radius=-0.25]
        arc[start angle=0, end angle=180, x radius=\rs, y radius=\rs];
        \clip (0,-\rs) rectangle (-\rs,\rs);
        \fill[draw=none, colS!50!black] (0,0) ellipse (0.3*\rs cm and \rs cm);
    \end{scope}

    \begin{scope}[shift={(\ls,0)}]
        \clip (-\rs,0) arc[start angle=180, end angle=0, x radius=\rs, y radius=-0.25]
        arc[start angle=0, end angle=180, x radius=\rs, y radius=\rs];
        \clip (90:\rn) arc (90:270:0.3*\rn cm and \rn cm) arc (-90:90:\rn);
        \fill[draw=none, ball color=colcore,shading=ball] (0,0) circle (\rn cm);
    \end{scope}

    \begin{scope}[shift={(\ls,0)}]
        \draw[snakearrow] (1,0) -- ++(0.7,0);
        \draw[snakearrow] (0.2,1) -- ++(0,0.7);
        \draw[snakearrow] (0.7,0.7) -- ++(0.55,0.55);
        \draw[snakearrow] (0.2,-1) -- ++(0,-0.7);
        \draw[snakearrow] (0.7,-0.7) -- ++(0.55,-0.55);
    \end{scope}

\end{tikzpicture}}
    \caption{\label{fig:five_planets}Jupiter, Saturn, Uranus, Neptune, and Earth with isotropic UV airglow (green) and polar aurorae (magenta), and internal heating (red arrows). Relative planetary sizes are to scale.}
\end{figure}

In this work, we broaden the framework to demonstrate the versatility of planetary energy-injection signals in probing the particle nature of DM, and emphasize the critical complementarity provided by different planetary systems. We focus on the interplay of indirect signals (class 2 above), as they are among the least model dependent; they require only that annihilation products deposit sufficient energy into the planet, and are otherwise insensitive to branching ratios or spectral distortions from mediator cascades~\cite{Elor:2015bho}. In particular, we focus on the interplay of two key observables, UV airglow and interior heating, schematically depicted in Fig.~\ref{fig:five_planets}. We do not consider \hp ionization since it applies only to Jupiter and Jupiter-like planets due to their unique atmospheric properties~\cite{Blanco:2023qgi}.

Interpreting these energy-injection signals requires understanding where and how DM deposits its annihilation products within the planet. In particular, the location and amount of energy deposition are determined by three key factors:
\begin{enumerate}
    \item the DM radial density and temperature distribution within the planet,
    \item the proximity of energy deposition by DM annihilation products with respect to the annihilation site, and
    \item the planet’s internal structure—specifically its SM radial distribution of density, temperature, as well as composition.    
\end{enumerate}
The DM distribution (1) and the displacement of annihilation products (2) are both governed by the underlying particle physics model. In particular, DM will settle into a different equilibrium distribution if it is lighter or heavier, as well as if it interacts strongly or weakly. Similarly, whether annihilation products are injected locally or at a displaced location is determined by the mass and lifetime of the mediator: a heavy or short-lived mediator decays near the annihilation site, while a light or long-lived mediator may transport energy over a macroscopic distance before decaying.

These considerations underscore the importance of planetary diversity for DM detection. Variations in planetary structure (3) alter where captured DM settles and annihilates, while differences in planetary radii determine which mediator decay lengths are most effectively probed. By taking all these effects into account, we demonstrate that the ensemble of Solar System planets, as well as large and dense Super-Jupiters, each with distinct properties, provides a uniquely powerful and complementary platform for exploring the dark sector.

This paper is organized as follows. In Section~\ref{sec:accumulation}, we review the accumulation of DM in planetary environments, detailing the capture process, annihilation, and evaporation, as well as the resulting radial distribution of DM. In Section~\ref{sec:overview}, we introduce the two indirect signatures investigated in this work — anomalous UV airglow and internal heat flow — and discuss how DM annihilation can modify these observables. Sections~\ref{sec:heavy_mediator} and \ref{sec:light_mediator} present results for two benchmark scenarios: DM annihilation via a heavy mediator (with respect to the center-of-mass energy), resulting in prompt energy deposition, and annihilation via a boosted or long-lived mediator, producing displaced energy injection. We compare the sensitivity of different planets in each case, highlighting their complementary reach. Finally, we summarize our findings and discuss future directions in Section~\ref{sec:summary}.

\section{\label{sec:accumulation}Dark Matter Accumulation in Planets}

DM can accumulate in planets by scattering with their constituents, losing energy, and becoming gravitationally captured. The accumulated DM particles may annihilate, producing detectable signals, and can come into an equilibrium between capture and annihilation. If DM is too light, it can escape the gravitational field of the object due to thermal kicks from the SM matter, $i.e.$, it can evaporate. The DM number $N_\chi$ within each planet can be obtained through the differential equation,
\begin{align}\label{eq:diff_eq_Nchi}
    \dv{N_\chi}{t} &=  C_\text{cap} - N_\chi^2 C_\text{ann} - N_\chi C_\text{evap} \ ,
\end{align}
where $C_\text{cap}$, $C_\text{ann}$, and $C_\text{evap}$ are the capture, annihilation, and evaporation rates respectively; we detail each contribution in the subsections below. 

\subsection{\label{sec:capture}Capture Rate}

The DM capture rate can be approximated as~\cite{Garani:2017jcj,Leane:2020wob}
\begin{align}\label{eq:C_cap}
    C_\text{cap} &= \pi R^2 \sqrt{\frac{8}{3\pi}} \frac{\rho_\chi}{m_\chi} v_\chi \left( 1 + \frac{3 v_\text{esc}^2}{2 v_\chi^2} \right)\times f_{\rm cap} \ ,
\end{align}
where ${\rho_\chi = 0.4}$~GeV/cm$^3$ is the local DM density, ${v_\chi \sim 270}$~km/s is the local DM velocity dispersion~\cite{McMillan:2009yr}, $m_\chi$ is the DM mass, $R$ is the planetary radius, ${v_\text{esc} = \sqrt{2G M/R}}$ is the escape velocity at the planetary surface for planetary mass $M$, where $G$ is the gravitational constant, and $f_\text{cap}$ is the fraction of captured DM particles, which scales with the scattering cross section and DM mass. For simplicity, we neglect the planet's motion relative to the DM halo, as it introduces a minor correction. We use \texttt{Asteria} to calculate a given planet's capture rate~\cite{Leane:2023woh,asteria}, which incorporates regimes where DM is captured after a single or multiple scatters, and includes the reflection of light DM.

The scattering rate for capture depends on the planetary composition, as detailed in Appendix~\ref{sec:properties}. For spin-independent interactions, the scattering cross section is
\begin{align}\label{eq:SI_cross_section}  
     \sigma_{\chi N}^\text{SI} &=  \frac{\mu_{\chi N}^2}{\mu_{\chi j}^2} A^2 \sigma_{\chi j}^\text{SI} \ , 
\end{align}
where $\mu_{\chi N}$ and $\mu_{\chi j}$ are the DM-nucleus and DM-nucleon reduced masses, respectively, and $A$ is the atomic mass number of the target nucleus.

Spin-dependent DM-SM scattering is given by
\begin{align}
    \sigma_{\chi N}^\text{SD} &=  \frac{\mu_{\chi N}^2}{\mu_{\chi j}^2} \frac{4(J+1)}{3J} \ev*{S_{\lbrace p,n \rbrace}}^2 \sigma_{\chi \lbrace p,n \rbrace}^\text{SD} \ ,
    \label{eq:SD_cross_section}
\end{align}
where $\ev*{S_{\lbrace p,n \rbrace}}$ is the expectation value of the spin of the proton or neutron within the nucleus~\cite{Bednyakov:2004xq} and $J$ is the spin of the nucleus. We consider the case where the proton interactions are turned on separately from the neutron, each under the assumption that the other is negligible.

\subsection{\label{sec:DMprofile}Radial Distribution}

Following capture, DM can thermalize with the planetary interior either locally or globally. The equilibrium distribution of DM is determined by the interplay between the gravitational potential, SM temperature, and the DM scattering cross section and mass. When the mean free path of DM is short compared to the planetary radius, corresponding to large cross sections, the DM remains in local thermal equilibrium (LTE) with the planet. In this regime, the DM temperature distribution is similar to the planet’s thermal profile. Conversely, when the mean free path is long, corresponding to small cross sections, DM particles interact infrequently, typically scattering once or less per planetary crossing. In this limit, the DM temperature distribution becomes approximately isothermal and no longer traces the internal temperature gradient of the planet. This temperature dependence is important for the density distribution for DM, which we now discuss.

\begin{figure*}[t]
    \centering
    \begin{minipage}[b]{0.5\linewidth}
    \centering
    \includegraphics[width=\linewidth]{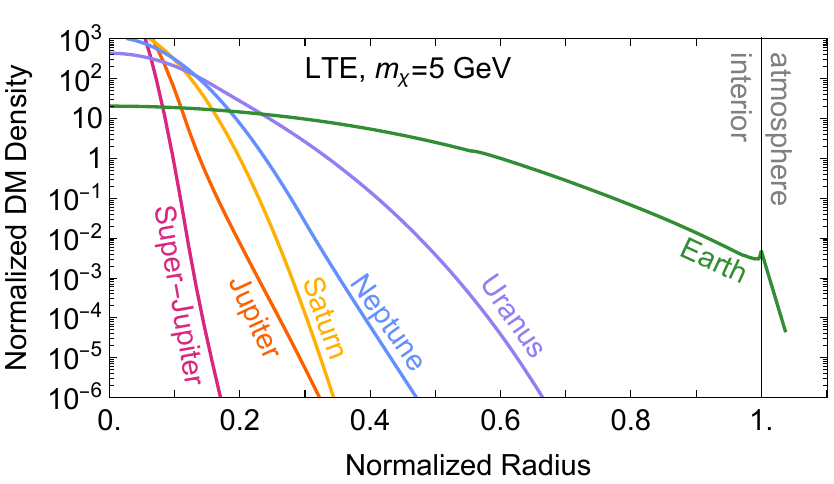}
  \end{minipage}
  \begin{minipage}[b]{0.5\linewidth}
    \centering
    \includegraphics[width=\linewidth]{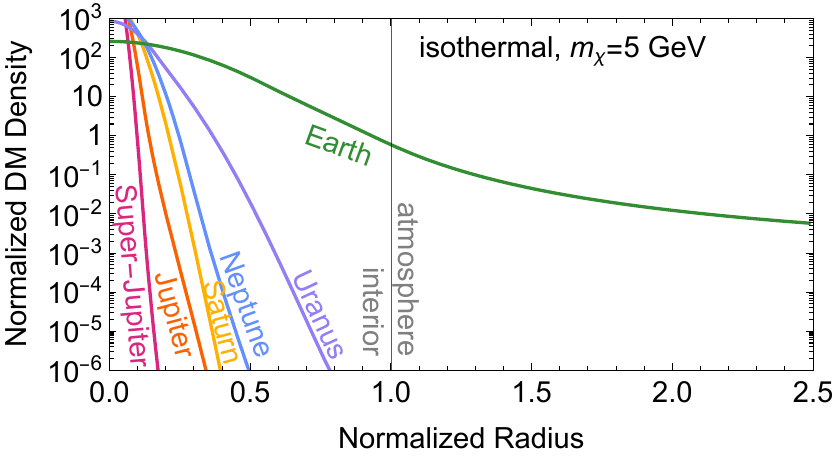}
  \end{minipage}

    \begin{minipage}[b]{0.5\linewidth}
    \centering
    \includegraphics[width=\linewidth]{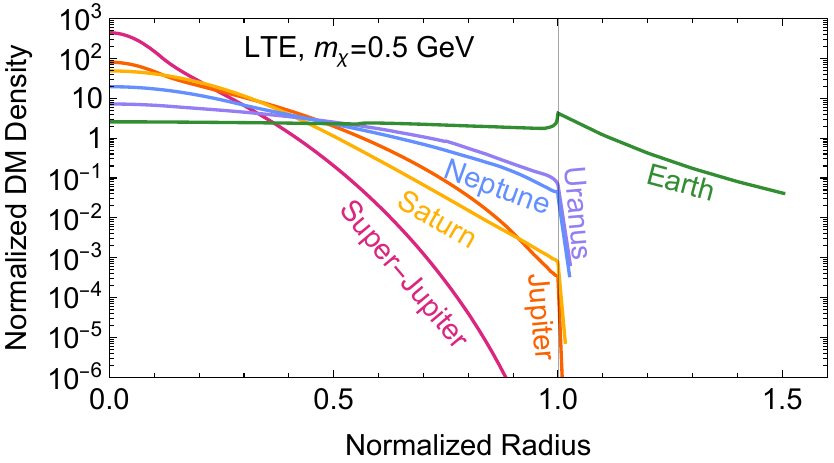}
  \end{minipage}
  \begin{minipage}[b]{0.5\linewidth}
    \centering
    \includegraphics[width=\linewidth]{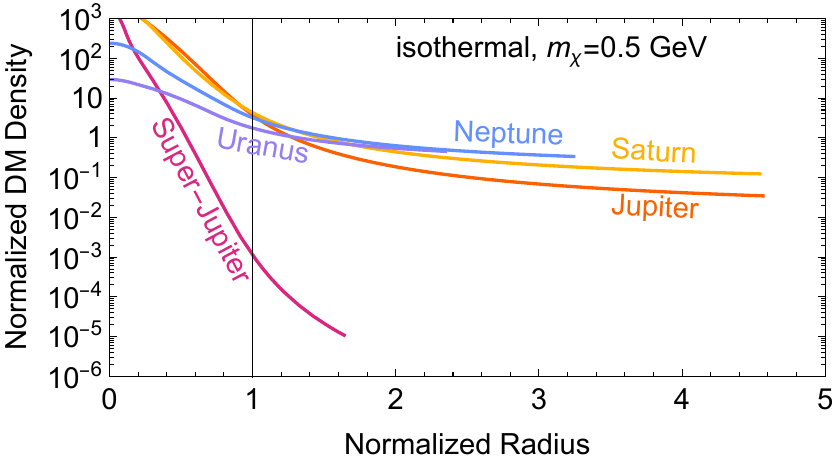}
  \end{minipage} 
    \caption{DM radial number density profile in the Solar System planets and a Super-Jupiter as labeled, normalized to the planets' volume, for DM with a mass of 5~GeV~(top) and 0.5~GeV~(bottom). We show profiles for DM that has a large cross section and so thermalizes with its local environment in the interior, called the local thermal equilibrium (LTE) regime~(left), and profiles for DM that has a small cross section and so gravitationally orbits many times between scatters and thermalizes with the planet as a whole, called the isothermal regime~(right). The distance to the planet cores is normalized by each planetary radius; values above 1 indicate the atmosphere, which have different extent for different objects; see text for details.}
    \label{fig:DMprofile}
\end{figure*}

\subsubsection{Long Mean Free Path Regime}

In the isothermal regime, the radial density profile $n_\chi^\text{iso} (r)$ is given by~\cite{Spergel:1984re, Scott:2008ns, Vincent:2015gqa, Acevedo:2023owd}
\begin{align} 
    n_\chi^\text{iso}(r) \propto e^{-U(r) / T_\chi^{\rm iso}} \ , 
\end{align}
where $\nabla U(r) = - m_\chi \vec{g}(r)$ is the gradient of the gravitational potential energy, with $g(r)$ the modulus of the gravitational acceleration at distance $r$ from the center of the planet. We take 
\begin{align}
    T_\chi^\text{iso} = \frac{T(0) + T(\min(r_\chi,R))}{2}
\end{align}
as the isothermal DM temperature, where $T(r)$ is the planet's temperature distribution, and $r_\chi$ is the DM scale radius, given by
\begin{align}\label{eq:rchi}
    r_\chi = \left( \frac{3 T(0)}{2\pi G \rho(0) m_\chi} \right)^{1/2} \ ,
\end{align}
where $\rho(r)$ is the planet's SM density. To encompass all scale radius scenarios, we take the minimum between the scale and planet radii in defining the temperature. The temperature and density profiles we use are outlined in Appendix~\ref{sec:properties}. {For simplicity, we have assumed the isothermal temperature is the average of the scale radius and core temperature. This is because the DM temperature is somewhere between the temperature at the scale radius and the core temperature upon closer analysis~\cite{Gould:1989ez, Banks:2021sba}, and in contrast to the original work of Ref.~\cite{Spergel:1984re}.}

\subsubsection{Short Mean Free Path Regime}

In the LTE regime, we determine the DM radial density profile $n_\chi^\text{LTE} (r)$ by solving the differential equation given by~\cite{Gould:1989hm, Vincent:2013lua, Banks:2021sba, Leane:2022hkk, Banks:2024eag}
\begin{align}\label{eq:DM_radial_profile}
    \dv{r} \log(n_\chi^\text{LTE}(r) T(r)^{\kappa + 1}) + \frac{m_\chi g(r)}{T(r)} &= 0 \ ,
\end{align}
where ${\kappa \sim - 1 / [2 (1 + m_\chi / m_\text{SM})^{3/2}]}$ is the diffusion coefficient~\cite{Leane:2022hkk}, with $m_\text{SM}$ being the SM target mass. For simplicity, we do not account for DM particles that have not yet fully thermalized and are in the process of slowing down within the planet~\cite{DeLuca:2018mzn, Pospelov:2019vuf, Pospelov:2020ktu, Rajendran:2020tmw, Budker:2021quh, McKeen:2022poo, Billard:2022cqd, Leane:2022hkk}, which would only enhance the DM signal.

\subsubsection{Treatment of the Dark Matter Atmosphere}

To model the radial distribution of DM in the atmosphere of the planets, we adopt an additional isothermal model. Since the SM density in the atmosphere drops severely, atmospheric DM is not in local equilibrium with the rarefied gas, even at the largest cross sections we consider. Indeed, for example, in Earth's planetary atmosphere it is well known that above a certain height, the density of SM gases is low enough and the temperature constant enough that the dynamics are best described by an isothermal layer, see $e.g.$ Refs.~\cite{Gold_isothermal, Jeans}.

For weak cross sections, the DM profile is already isothermal, and so the treatment above is sufficient as is. However, for large cross sections where the DM is in LTE inside a given planet, its radial distribution is taken to be an isotherm in the atmosphere with a temperature that is set to be the temperature at the surface of the planet. The modified atmospheric distribution for the interior LTE and atmospheric isotherm is taken to be
\begin{align}
    n_\chi(r) &=  
  \begin{cases}
    n_{\chi}^\text{LTE} (r) \ , &  r \leq R \\
    n_{\chi}^\text{LTE} (R) \times e^{-(U(r)-U(R))/T(R)} \ ,               & r >  R
  \end{cases} \ .
\end{align}

There is the additional question as to where the edge of the DM atmosphere extends.
Accumulated DM particles can temporarily be located outside of the planet, and even outside of its atmosphere, while remaining gravitationally bound. We account for these particles in the radial distribution and define a mass-dependent maximal radius,
\begin{equation}
    R_\text{max} \simeq \frac{G M m_\chi}{ T_\chi} \ ,
\end{equation}
where the thermal velocity $v_\text{th}$ becomes larger than the escape velocity. We take $T_\chi=T_\chi^\text{iso}$ for the isothermal regime and $T_\chi=T(R)$ for LTE. In practice, we truncate the curves at a radius of either $R_\text{max}$, or once the DM density drops two orders of magnitude with respect to the DM surface density, as it is more numerically stable and any smaller DM density contributes negligibly to the rate.

Instead of taking the treatment above, it is also possible to simply truncate the DM profiles once the SM profile data ends. In our parameter space which yields observable signals, we find such an approach provides comparable (within $\sim10\%$) results. 

\subsubsection{Comparison of Regimes}

To compare the DM distribution across different planets and DM masses, we define the normalized DM radial profile $G_\chi (r)$ by~\cite{Pospelov:2023mlz} 
\begin{align}
    G_\chi(r) \equiv \frac{n_\chi(r) V_\text{max}}{4 \pi \int_0^{R_\text{max}} \dd{r'} r'^2 n_\chi(r')} \ ,
\end{align}
where $V_\text{max}$ is the planet's volume up to $R_\text{max}$.

Figure~\ref{fig:DMprofile} shows the normalized LTE and isothermal DM radial profiles, assuming DM masses of 0.5 and 5~GeV. The lighter the DM mass, the flatter the profile, with further profile extent towards the atmosphere. Heavy DM, on the other hand, tends to concentrate near the core. For the LTE profile, the peak observed at the boundary between the interior and atmosphere is due to the temperature reaching a minimum while the density of atomic elements remains relatively large, enhancing the probability that DM remains in a trough between the hotter interior and upper atmosphere. The isothermal distribution on the other hand does not depend on the local temperature, and therefore is not enhanced at the interior-atmosphere boundary. We define the interior-atmosphere boundary $R$ to be the planet radius where the pressure drops to 1~bar. In Fig.~\ref{fig:DMprofile}, Earth is not shown for the 0.5 GeV DM mass isothermal case, as it is not well-defined, with a scale radius extending outside the planet; in practice this parameter space is never reached due to evaporation. {For each of the planets, the transition between the LTE and isothermal regimes occurs around approximately $\lbrace 1, 3, 50, 40, 60 , 0.1 \rbrace \times 10^{-34}$~cm$^2$ for Jupiter, Saturn, Uranus, Neptune, Earth, and Super Jupiter, respectively. For a given planet, we therefore refer to ``large" cross sections as above this number, and ``small" as below.}

\subsubsection{Transition Between Regimes}

The results for the regimes above apply to the limit of large or small cross sections. 

For intermediate cross sections, the profile is a combination of the two. In this transition regime, we combine these distributions by taking~\cite{Scott:2008ns}
\begin{align}
    n_\chi(r) &= \mathfrak{f}(K) n_\chi^\text{LTE} (r) + \big[ 1-\mathfrak{f}(K) \big]n_\chi^\text{iso} (r) \ ,
\end{align}
with 
\begin{align}
    \mathfrak{f}(K)=1-\left(1+\left(\frac{K_0}{K}\right)^{1/\varsigma}\right)^{-1} \ ,
\end{align}
where $K_0 = 0.4$ and $\varsigma = 0.5$ are free best-fit parameters based on simulations in the Sun~\cite{Scott:2008ns}; similar results have been used for other objects~\cite{Banks:2024eag}.
This blends the distributions of the two limiting regimes depending on the Knudsen number, $K$, a ratio of the mean free path to the scale radius, which is given by 
\begin{align}
    K = \frac{\lambda_\chi(0)}{r_\chi} \ ,
\end{align}
where the mean free path of the DM is
\begin{align}
	\lambda_\chi(r) = \left( \sum_N n_N(r) \sigma_{\chi N} \right)^{-1} \ ,
\end{align}
where $n_N(r)$ is the SM number density at radius $r$.

\subsubsection{Advective Effects}

When the cross section between DM and SM particles is very large, the DM profile may also be affected by advection. To determine when this is relevant, we compare advection with the diffusion timescale. For light DM, we take the diffusion timescale to be~\cite{Leane:2022hkk}
\begin{align}
    \tau_\text{dif} \simeq \frac{R^2 n_\text{SM} \sigma_{\chi N}}{v_\text{th}} \ ,
\end{align}
where we use the volume-averaged planet temperature and density, and $m_\text{SM}$ from Table~\ref{tab:planet_data}. We assume the advection timescale $\tau_\text{adv}$ is 100~years in the interior of giant planets, and one million years in the Earth~\cite{Leane:2022hkk}. For the DM profile calculation above to remain unimpacted, we require $\tau_\text{dif} \lesssim \tau_\text{adv}$. Over the DM mass range we consider, we find that cross sections below $10^{-28}$~cm$^2$ in the giant planets and $10^{-24}$~cm$^2$ in the Earth satisfy this constraint. For larger cross sections, we do not implement advection, but its effect should be kept in mind as an uncertainty in such regions. 

While advective timescales can be much faster than quoted above in the atmospheres, {this does not significantly perturb the DM distribution. The atmospheric gas iself follows an adiabatic temperature profile, but the DM population maintains an approximately isothermal equilibrium profile, characterized by the local temperature. As long as DM's} mean free path is not shorter than the scale height of the atmosphere, which is satisfied for our parameter space, {advection in the atmosphere does not appreciably modify the DM density profile}.

\subsection{\label{sec:equilibrium}Annihilation Rate and Equilibrium with Capture}

Once the DM is captured and thermalizes into the radial profile, it can annihilate. As per our companion paper, we use~\cite{Blanco:2024lqw}
\begin{align}\label{eq:Gamma_vs_C}
    \Gamma_\text{ann} = m_\chi N_\chi^2 C_\text{ann} \ ,
\end{align}
where the annihilation rate is given by
\begin{align} \label{eq:C_ann}
    C_\text{ann} &= \frac{1}{V_\text{max}^2} \int_{V_\text{max}} \dd{V_\text{max}} \ev{\sigma v} G_\chi^2(r) \ ,
\end{align}
and $\ev{\sigma v}$ is the thermally averaged annihilation cross section.

For detectable annihilation signals, DM evaporation must be negligible. In this case, the solution to the differential equation in Eq.~(\ref{eq:diff_eq_Nchi}) is given by
\begin{align}\label{eq:NDM}
    N_\chi (t) &= \left( \frac{C_\text{cap}}{C_\text{ann}} \right)^{1/2} \tanh(\frac{t}{t_\text{eq}}) \ ,
\end{align}
where the characteristic equilibration timescale for capture and annihilation $t_\text{eq}$ is given by
\begin{align} \label{eq:t_eq}
    t_\text{eq} &\equiv \frac{1}{\sqrt{C_\text{cap} C_\text{ann}}} \ .
\end{align}
After equilibration between the capture and annihilation rate, the number of accumulated DM particles is constant and given by
\begin{align}
\label{eq:ann_num}
    N_\chi (t \gtrsim t_\text{eq}) = \left( \frac{C_\text{cap}}{C_\text{ann}} \right)^{1/2} \ .
\end{align}

We assume capture-annihilation equilibrium. This can be verified directly by ensuring that the equilibrium timescale, defined in Eq.~\eqref{eq:t_eq}, is shorter than the age of the planet. We find that equilibrium is reached in the majority of our parameter space for the case of $s$-wave annihilation. It is also largely reached for $p$-wave annihilation for signals in the Super-Jupiter, but only for somewhat larger cross sections for local planets. We discuss this further in Appendix~\ref{sec:wave}.

For large thermally averaged annihilation rates, the DM radial density profile can also change compared to the equilibrium calculation performed above. Using Eq.~(\ref{eq:ann_num}) to determine a DM annihilation timescale, and comparing with the advection and diffusion timescales listed above, we have checked that for both $s$-wave and $p$-wave annihilation rates the equilibrium density distribution is not impacted in our considered parameter space.

\subsection{\label{sec:evaporation}Evaporation Rate}

In the process of DM evaporation, DM particles with sufficient kinetic energy escape the planet's gravitational potential. As with capture, we assume contact interactions between the DM and SM (long-range interactions instead can yield different results~\cite{Acevedo:2023owd}). We adopt a DM evaporation rate given by~\cite{1990ApJ...356..302G,Acevedo:2023owd, Garani:2021feo}
\begin{align} \label{eq:C_evap_gould}
    C_{\text {evap}}= \frac{4 \pi}{{V_\text{max}}} \int_0^{{R_\text{max}}} \dd{r} r^2 s(r) G_\chi(r) \xi(r) \lambda_\chi^{-1} (r) \ .
\end{align}
Here $\xi$ is the evaporation function, which is a velocity-weighted measure of the number of DM particles with kinetic energy exceeding the local gravitational potential energy, and is given by
\begin{align}
    \xi(r)=\bar{v}(r)\left|\frac{U(r)}{T(r)}\right| e^{-\left|U(r)/T(r) \right|} \ ,
\end{align}
with $\bar{v}=\sqrt{8 T(r) /\left(\pi m_\chi\right)}$. The approximate escape probability is
\begin{align}
    s(r)= \frac{7}{10} \exp \left[  \eta(r) \left( \frac{3\, T(r)}{2 \, U(r)} - 1\right)  \right]\min \left(1, \eta^{-1}(r)\right) \ ,
\end{align}
and $\eta$ is the opacity given by 
\begin{align}
    \eta(r)=\int_r^{{R_\text{max}}} \dd{r'} \lambda_\chi^{-1}(r') \ .
\end{align}
DM evaporates rapidly from the planet when the evaporation rate in Eq.~\eqref{eq:C_evap_gould} satisfies
\begin{align}\label{eq:evap_condition}
C_\text{evap} > \sqrt{C_\text{cap} C_\text{ann}} \ .    
\end{align}
Once the DM mass is sufficiently light the evaporation rate rapidly increases. Therefore, this condition serves as an approximate ``cut-off" for low DM-mass sensitivity.

Eq.~\eqref{eq:C_evap_gould} gives the full evaporation rate valid for all cross sections where the planet's entire volume contributes to the final rate. However, for increasingly large cross sections, the most likely evaporation shell becomes increasingly thin and rises towards the surface, making Eq.~\eqref{eq:C_evap_gould} computationally inefficient. Therefore, at large cross sections ($i.e.$, above the object's transition cross section), we use the Jeans framework, which instead assumes that DM always evaporates from the surface of last scattering, one mean free path below $R$. We adopt the Jeans evaporation rate~\cite{Jeans, Neufeld:2018slx, Moore_2021}  
\begin{align}\label{eq:C_evap_jeans}
    C_\text{evap}^\text{J} &= \frac{\Phi_J S}{N_\chi} \ , 
\end{align}
where $S$ is the surface area of the planet. The Jeans flux $\Phi_J$ given by
\begin{align}
    \Phi_J &= \frac{n_\chi(r_\text{LSS}) v_\text{LSS}}{2 \pi^{1/2}} \left( 1 + \frac{v_\text{esc}^2}{v_\text{LSS}^2} \right) e^{-v_\text{esc}^2/v_\text{LSS}^2} \ ,  
\end{align}
where $r_\text{LSS}$ is the radius of last scattering given by
\begin{align}
    \eta(r_\text{LSS}) &= 1 \ ,
\end{align}
and $v_\text{LSS} = \sqrt{2 T_\text{LSS} / m_\chi}$ is the most probable thermal velocity at the last scattering surface. We enforce a maximum $r_{\text{LSS}}$ equal to $R$, the radius of the object, since we assume that the DM becomes isothermal in the atmosphere in the large cross section regime.
Finally, we set the evaporation mass limit for Eq.~\eqref{eq:C_evap_jeans} using the same condition in Eq.~\eqref{eq:evap_condition}.

\section{\label{sec:overview} Planetary Spectroscopy and Energy Deposition Signals}

We study two indirect signatures of energy deposition in Solar System planets and Super-Jupiters: UV airglow from atmospheric excitation and excess internal heat, both detectable via planetary spectroscopy. Table~\ref{tab:airglow_and_heat} summarizes current measurements for UV airglow and internal heat across Solar System planets and projections for a benchmark Super-Jupiter. We first review known contributions to these observables and the datasets across planet classes, then assess potential signatures from DM annihilation.

Our analysis assumes DM annihilates into visible SM particles, $i.e.$, those that deposit energy before escaping the planet, such as charged leptons or hadrons, and neglects neutrinos. If some DM annihilation is to neutrinos, results can be rescaled by the visible branching ratio.

\begin{table*}[t]
    \caption{\label{tab:airglow_and_heat}UV night airglow/electron precipitation and internal heat flux of the Solar System gas and ice giants, as well as Earth and a hypothetical massive Super-Jupiter.}
    \centering
    \begin{tabular}{p{2.2cm} p{3.0cm} p{3.0cm} p{1.2cm} p{3.0cm} p{3.0cm} p{1.2cm}}
        \toprule 
        & \multicolumn{3}{c}{UV Airglow} & \multicolumn{3}{c}{Internal Heat} \\ \cmidrule(lr){2-4}\cmidrule(lr){5-7}
        & $P_\text{atm}^\text{observed}$ ($\mu$W/m$^2$) & Probe & Ref. & $P_\text{int}^\text{observed}$ (W/m$^2$) & Probe & Ref. \\ \midrule 
        Earth & $\phantom{<~}0.05$ & AMS-02, ELFIN& \cite{2024GeoRL..5105134Q,AMS:2000kgm,AMS:2014gdf, graziani2015time} & 0.092(4) & Boreholes & \cite{2010SolE....1....5D} \\
        Jupiter & $\phantom{<~}0.31_{-0.15}^{+0.19}$ & New Horizons & \cite{2011epsc.conf.1474B} 
        & 7.485(160) & Cassini & \cite{2018NatCo...9.3709L} \\
        Saturn & $< 1$ & Voyager 1 & \cite{1981Sci...212..206B}
        & 2.84(20) & Cassini & \cite{2024NatCo..15.5045W} \\
        Uranus & $\phantom{<~}4.6$ & Voyager 2 & \cite{1989Sci...246.1459B} & 0.042(47) & Voyager 2 & \cite{1990Icar...84...12P} \\
        Neptune & $\phantom{<~}1.9 \pm 0.3$ & Voyager 2 & \cite{1989Sci...246.1459B} & 0.433(46) & Voyager 2 & \cite{1991JGR....9618921P} \\
        \midrule
        Super-Jupiter & $<1.25$ & CR background & \cite{2018AA...614A.111P} & 459 & \texttt{ATMO} model & \cite{2020AA...637A..38P} \\
        \bottomrule
    \end{tabular} 
\end{table*}

\subsection{Review of UV Airglow and Datasets}

On the dayside of planetary atmospheres, high-energy solar radiation ionizes atoms and molecules. The resulting electrons and ions precipitate and excite atmospheric species, producing UV emission known as UV dayglow. Although both ions and electrons can contribute, the UV emission is primarily driven by electrons~\cite{2016Icar..268..215G}. After dusk, the drop in solar ionizing flux allows charged particles to neutralize or deplete, causing the UV nightglow to be several orders of magnitude dimmer. These emissions, collectively termed airglow, provide a sensitive probe of energy deposition in planetary atmospheres and can reveal non-standard energy sources, such as DM annihilation.

We define the edge of the region of the atmosphere where energy deposition can produce UV airglow by the exobase height $h_\text{exo}$, measured from the planetary surface. The exobase marks the lower boundary of the exosphere—the outermost atmospheric layer—with values provided in Appendix~\ref{sec:properties}. The corresponding atmospheric radius is then $R_\text{atm} = R + h_\text{exo}$.

\subsubsection{Gas and Ice Giants}

Space probes such as Voyager~1, Voyager~2, and New Horizons have obtained UV nightglow spectra of the gas and ice giant planets away from their polar auroral regions. Our analysis of the Solar System giant planets focuses on the excitation of the Lyman and Werner bands of molecular hydrogen~(H$_2$), which dominate far UV emissions and contain minimal background contamination within the Solar System. The contribution from the Lyman and Werner bands has been extracted and converted into a physical brightness in rayleighs~(R) by comparing these observed spectra with laboratory reference H$_2$ spectra. 

To convert the UV airglow brightness into an equivalent precipitating power, we adopt a commonly used conversion factor of approximately 10~R per 1~$\mu$W/m$^2$ for electrons~\cite{1982JGR....87.4525G, 1983JGR....88.6143W, 2012JGRA..117.7316G}. More recent studies indicate a weak correlation between brightness and mean precipitating electron energy, though the relationship is highly model-dependent. For instance, an electron energy flux of 1~$\mu$W/m$^2$ could correspond to 14.6~R for 20~keV electrons or 7.8~R for 200~keV electrons~\cite{2016Icar..268..215G}. Given the moderate, model-dependent brightness changes, we adopt the conversion factor of 10~R per $\mu$W/m$^2$ of equivalent ionizing power. 

\subsubsection{Earth}

Earth’s atmosphere differs markedly from those of the giant planets due to its much lower hydrogen content. Nonetheless, a UV nightglow has been observed, $e.g.$ by the Universitetskii-Tatiana satellite~\cite{2011P\string&SS...59..733D}. Thanks to the wealth of experiments in the Earth's upper atmosphere, the precipitating electron energy flux can be directly measured, for example using CubeSats~\cite{2024GeoRL..5105134Q} or the AMS-02 experiment onboard the International Space Station located in low orbit~\cite{AMS:2000kgm, AMS:2000qrn, Aguilar:2013qda, AMS:2014gdf, Accardo:2014lma, AMS:2014xys, Consolandi:2014uia, aguilar2015precision, graziani2015time}. Moreover, both CubeSats and AMS-02 are able to distinguish electrons from protons~\cite{AMS:2000kgm, 2024GeoRL..5105134Q}, and AMS-02 can measure the positron spectrum independently from that of the electrons, $e.g.$~\cite{Aguilar:2013qda}. 

The fluxes and spectra of upward and downward electrons and positrons are nearly identical~\cite{AMS:2000kgm}. We model the equivalent ionizing power by considering upward-going positrons, which are the closest analogs of DM annihilation products, $i.e.$, annihilation products should be globally charge-neutral and should contribute half of their energy to the positron spectrum.  The ionizing upward-going electron spectrum is modeled by extending the spectrum observed by the ELFIN CubeSats (which only cover the range 0.05 to 5~MeV) via an interpolating power law to match the AMS-02 data to cover the range 30~keV up to 100~GeV. This results in a total electron ionization power of $\sim 0.5~\mu$W/m$^2$. Since the positron to electron ratio is $\sim 10\%$ across the energy range~\cite{Aguilar:2013qda, AMS:2014xys, Accardo:2014lma}, the maximum contribution of any additional source of equal-parts positrons and electrons is 0.05~$\mu$W/m$^2$.

\subsubsection{Local Free-Floating Super-Jupiter}

To investigate an exoplanetary scenario, we use a benchmark free-floating Super-Jupiter; free-floating targets will be optimal due to their lower backgrounds. The main origin of UV airglow in this case is not stellar, but rather from cosmic-ray electrons precipitating in the atmosphere. We calculate the expected electron energy flux with energies above 15.4~eV (the ionization energy of H$_2$) for a local free-floating Super-Jupiter by integrating the interstellar cosmic ray-electron spectrum from Ref.~\cite{Padovani:2018zrg}. We choose a local Super-Jupiter, rather than a distant one in a higher DM density environment, as UV airglow is difficult to measure at large distances due to dust extinction.

\subsection{Dark Matter Induced UV Airglow}

As pointed out in our companion paper~\cite{Blanco:2024lqw}, precipitating electrons from DM annihilation can generate anomalous UV airglow in planetary atmospheres. To quantify the effective DM-induced power injected into the atmosphere, we express the precipitating power from DM annihilation as
\begin{align}
   P_\text{atm}^\text{DM} &= \frac{\Gamma_\text{ann} \times \fatm {\times f_\text{dep}}}{4 \pi R^2} \ ,
    \label{eq:dmpower}
\end{align}
where $\fatm$ is the fraction of annihilation events injecting energy directly into the atmosphere, $\Gamma_\text{ann}$ is the annihilation rate defined in Eq.~\eqref{eq:Gamma_vs_C}{, and $f_\text{dep}$ is the fraction of the injected energy that is deposited through collisional processes capable of exciting atmospheric molecules}. {In contrast with our companion paper~\cite{Blanco:2024lqw}, here we do not assume ${f_\text{atm} \sim 1}$, and instead explicitly compute it for a wide range of models.} This $\fatm$ parameter depends on the type and energy of particles produced by DM annihilation. We focus on the energy of annihilation events rather than the specific energy spectrum of the precipitating particles, as the excitation threshold is orders of magnitude lower than the peak of the annihilation spectrum, such that effectively all of the annihilation energy is available to excite the molecules. 

To constrain the DM-induced airglow power $P_\text{atm}^\text{DM}$, Eq.~\eqref{eq:dmpower} can be compared directly to the equivalent ionizing precipitated power derived from the UV airglow brightness observed in the giant planets and Earth. We also set projected sensitivities for the case of a nearby free-floating Super-Jupiter which would be irradiated by cosmic-ray electrons. Using the power derived from direct observations $P_\text{atm}^\text{observed}$ presented in Table~\ref{tab:airglow_and_heat}, we impose the condition
\begin{align}\label{eq:Power_airglow}
    P_\text{atm}^\text{DM} \leq P_\text{atm}^\text{observed} \ .
\end{align}
In comparing with the observed powers from Table~\ref{tab:airglow_and_heat}, our assumption is that the relationship between emission brightness and power injection is the same for both DM annihilation and astrophysical electron precipitation.

{For Earth, we set ${f_\text{dep} = 1}$, as the comparison is performed using direct measurements of precipitating electrons and positrons. For the giant planets and the Super-Jupiter case, the annihilation products are not directly observed and instead lead to to a UV airglow, and we therefore evaluate $f_\text{dep}$ as a function of the annihilating DM mass using the stopping power of electron in hydrogen using the ESTAR database from NIST~\cite{ESTAR}, the main constituent of the giant planets' atmosphere.}

\subsection{Review of Internal Heat Flow and Datasets}

Internal heat in planets can be indirectly measured using the planet's infrared~(IR) spectrum. This measurement could be sensitive to new sources of energy deposition, such as DM annihilation products.

\subsubsection{Gas and Ice Giants}

In the giant planets, the internal heating is estimated by evaluating the photon power absorbed by the planet. This is done by considering the photon power incident on the planet, and measuring the planet's Bond albedo (the fraction of scattered light), using optical instruments. The effective temperature of the planet is obtained via infrared measurement, and the net internal power is given by the difference of the observed effective power and the incident solar power~\cite{1980RvGSP..18....1H}.

The first accurate estimates of the gas giants internal heat were done by Voyager~1 and~2~\cite{1981JGR....86.8705H, 1983Icar...53..262H,  1990Icar...84...12P, 1991JGR....9618921P}. More recently, the Cassini mission provided higher-precision data for Jupiter and Saturn, benefiting from improved instrumentation and a wider range of observation angles and wavelengths~\cite{2018NatCo...9.3709L, 2024NatCo..15.5045W}.

\subsubsection{Earth}

On Earth, several techniques are used to measure the internal heat. Oceanic and continental boreholes are used to directly measure the temperature without the influence of variation in the Earth’s surface temperature, from $e.g.$ day-night variance~\cite{2005RSPTA.363.2777H}. We use recent results which take into account effects from the local environment, $i.e.$, matching the local geology to the heat flow measurement, which incorporates heat absorption of local rocks~\cite{2010SolE....1....5D}. 

\subsubsection{Super-Jupiter in the Galactic Center}

We adopt the internal heat flux of a 10~Gyr Super-Jupiter from the \texttt{ATMO} model~\cite{2020AA...637A..38P}. We take this benchmark to be free-floating at 0.1~kpc from the Galactic Center, such that it is not irradiated by a star and its internal heat is a remnant from its formation.

\subsection{DM Induced Internal Heat Flow}

DM annihilation can inject energetic particles into planetary interiors which can very quickly deposit energy in their surroundings, generating anomalous heating. The total DM-induced internal heat flow power is
\begin{align}
   P_\text{int}^\text{DM} &= \frac{\Gamma_\text{ann} \times \fint}{4 \pi R^2} \ ,
    \label{eq:DM_heating}
\end{align}
where $\fint$ is the fraction of annihilation events that deposit energy into the planetary interior. Since modeling the detailed interior dynamics of planets is complex, we set conservative calorimetric limits by requiring that DM annihilation inside the planet must not contribute more thermal energy than what is observed flowing out of the planet's surface. This gives the constraint
\begin{align}\label{eq:Power_heat_flow}
    P_\text{int}^\text{DM} \leq P_\text{int}^\text{observed} \ ,
\end{align}
where $P_\text{int}^\text{observed}$ is the observed internal heat flow from the planet, presented in Table~\ref{tab:airglow_and_heat}. If DM were to produce more heat than what is observed, it would result in anomalous internal heating, offering a detectable signature of DM interactions within planetary interiors. Recent work also considered the impact of DM heating on Earth's core itself~\cite{Cappiello:2025yfe}, which we conservatively neglect.

We also investigate the DM-induced anomalous heating of a free-floating Super-Jupiter located at 0.1~kpc from the Galactic Center. We estimate the DM density at this location by making use of the generalized Navarro-Frenk-White profile~\cite{Zhao:1995cp} with ${\gamma = 1.5}$, resulting in an enhancement of the DM density with respect to the local value by a factor of ${\sim 1500}$. We conservatively use the same DM velocity $v_\chi$ as in the Solar neighborhood, as there are large uncertainties in the velocity near the Galactic Center. It is likely lower, and a smaller DM velocity would enhance the capture and annihilation rates, and increase the heat flow signal. 

We expect such a candidate to be detectable with telescopes such as JWST, see Ref.~\cite{Leane:2020wob} for detailed discussion of exposure times, instrument filters, etc.

When considering DM-induced planetary heat flow, we neglect contributions from dark kinetic heating~\cite{Acevedo:2024zkg}. These are expected to be negligible for the contact interactions we consider.

\section{\label{sec:heavy_mediator}Heavy Mediator Results}

We begin with the scenario in which DM annihilation products deposit their energy near the annihilation site. This is expected when the mediators are sufficiently heavy or short-lived. For simplicity, we refer to this as the “heavy mediator” scenario, which is a common assumption in contexts such as direct detection experiments. By determining the fraction of energy deposited in the atmosphere or interior, we can constrain the DM scattering cross section, or project sensitivities in the case of a Super-Jupiter, using UV airglow or internal heating as observational probes.

\subsection{Interplay of Internal vs Atmospheric Energy Deposition}

The fraction of DM annihilation events injecting energy into the atmosphere is given by
\begin{align}
    \fatm^\text{heavy} &= \frac{\int_R^{R_\text{atm}} \dd{r} r^2 G_\chi^2(r)}{\int_0^{R_\text{max}} \dd{r} r^2 G_\chi^2(r)} \ ,
\end{align}
where $G_\chi(r)$ is the normalized DM density profile, defined earlier in Section~\ref{sec:DMprofile}. The fraction of DM annihilation events that deposit energy into the planetary interior is then approximately
\begin{align}\label{eq:fint_fatm_heavy}
    \fint^\text{heavy} \simeq 1 - \fatm^\text{heavy} \ .
\end{align}

\begin{figure}[t]
    \centering
    \includegraphics[width=0.9\columnwidth]{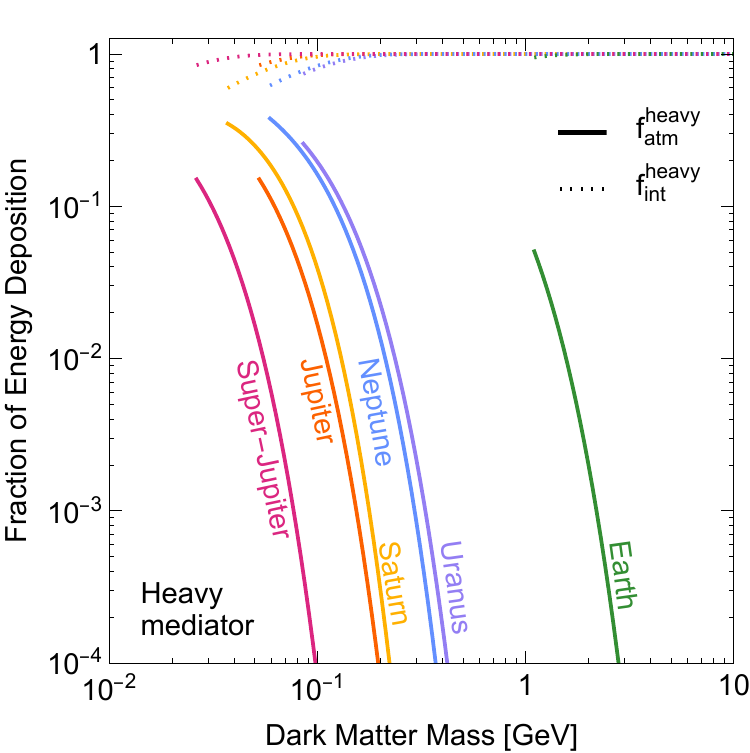}
    \caption{Fraction of energy deposited in the atmosphere ($\fatm^\text{heavy}$, solid) and interior ($\fint^\text{heavy}$, dotted) for DM annihilating through a heavy mediator, with DM distributed throughout the planet in the interior LTE regime, in the case of Jupiter, Saturn, Uranus, Neptune, Earth, and a Super-Jupiter.}
    \label{fig:fatm_fint}
\end{figure}

Figure~\ref{fig:fatm_fint} shows the fraction of energy deposited in the atmosphere ($\fatm^\text{heavy}$, solid lines) and in the interior ($\fint^\text{heavy}$, dotted lines) for DM annihilation through a heavy mediator, assuming the DM follows a local thermal equilibrium~(LTE) distribution in the interior of the planet. Results are shown for Jupiter, Saturn, Uranus, Neptune, Earth, and the benchmark Super-Jupiter. We have cut our results at the lowest DM mass that does not evaporate for the given planet. At lighter DM masses, more of the DM density profile extends into the atmosphere, and so lighter masses have the lightest atmospheric energy deposition. This shows that for light DM, even with a heavy mediator, a substantial atmospheric signal is possible. As the DM mass increases further, DM increasingly resides in the core. 

In contrast to $\fatm^\text{heavy}$, the fraction $\fint^\text{heavy}$ remains substantial across the entire mass range, reflecting the dominance of the planetary interior's volume. While these fractions are higher, they do not generally lead to superior heating limits compared to UV airglow, as the SM backgrounds from UV airglow are much lower.

\subsection{Cross Section Constraints and Projections}

Applying the atmospheric and interior energy deposition fractions, $\fatm^\text{heavy}$ and $\fint^\text{heavy}$, as well as accounting for the evaporation mass cutoff, we derive constraints and projections on the DM scattering cross section parameter space using anomalous UV airglow and internal heat flow limits, under the heavy mediator assumption.

\begin{figure*}[ttt]
    \centering
    \includegraphics[width=0.44\textwidth]{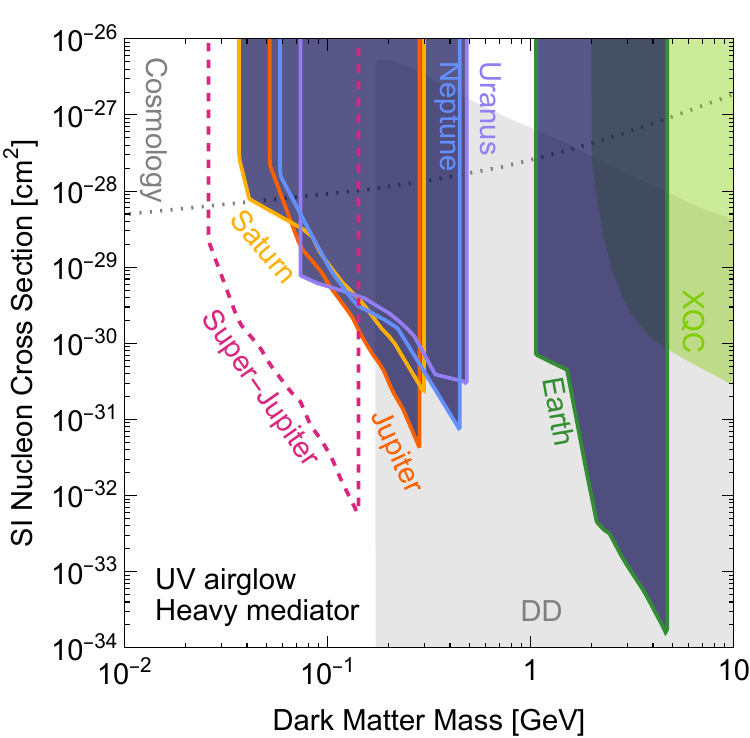}\hspace{4mm}
    \includegraphics[width=0.44\textwidth]{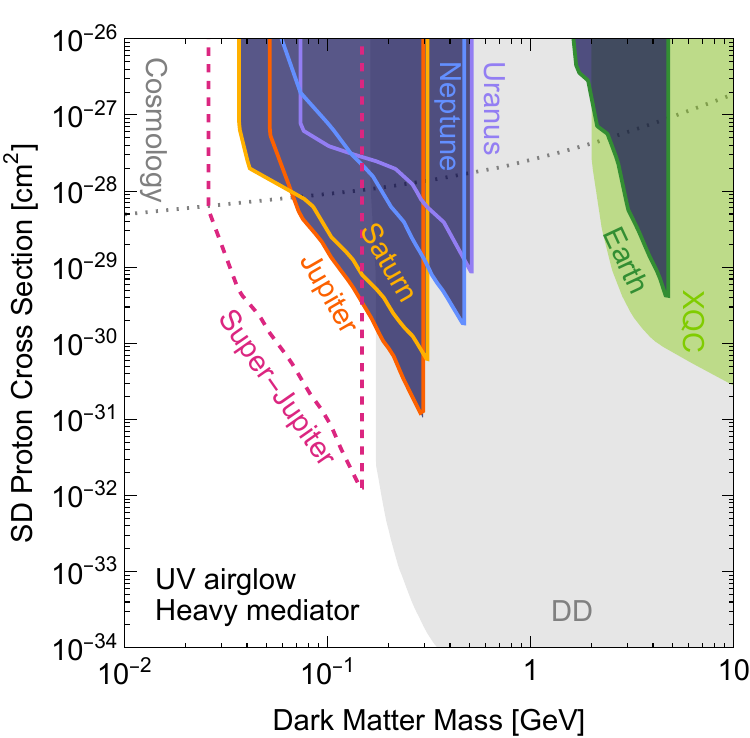}
    \includegraphics[width=0.44\textwidth]{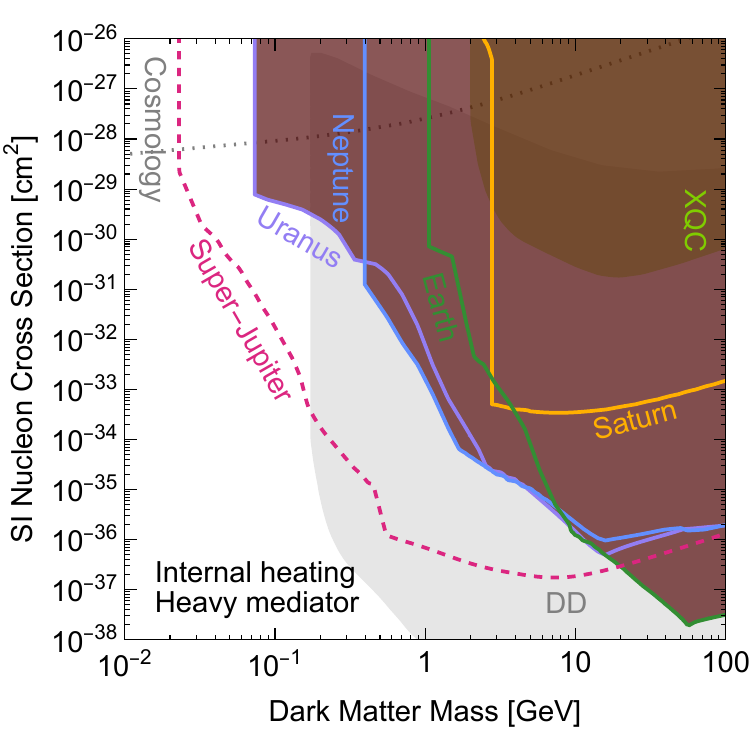}\hspace{4mm}
    \includegraphics[width=0.44\textwidth]{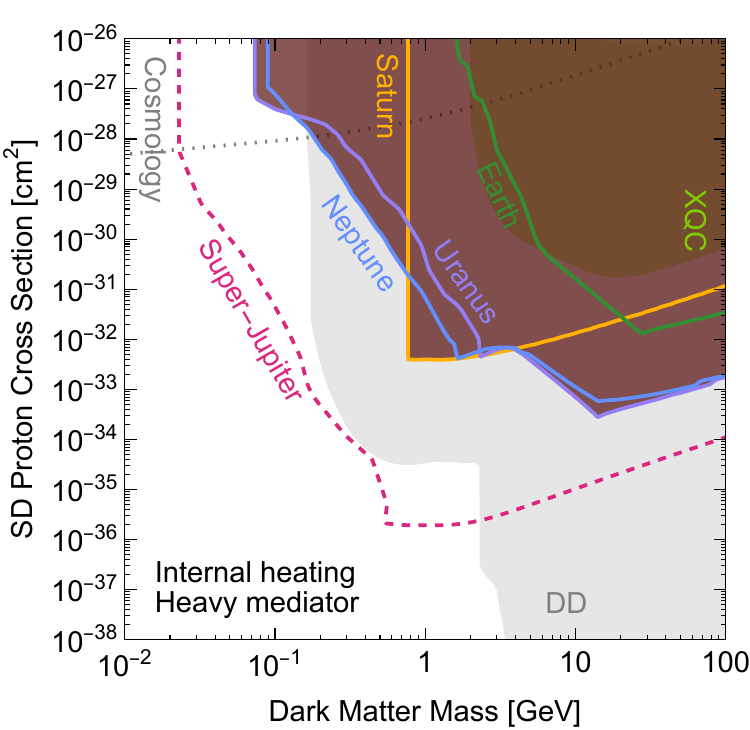}
    \caption{\label{fig:heavy_mediator}
    Constraints (solid, shaded) and future projections (dashed) on spin-independent (SI) nucleon scattering (left) and spin-dependent (SD) proton scattering (right), assuming DM annihilation through a heavy mediator using planetary UV airglow (top) and internal heat flow (bottom).  The gray region is excluded by direct detection (DD) experiments~\cite{PICO:2019vsc, CRESST:2019jnq, CRESST:2022dtl, DarkSide:2022dhx, LZ:2024zvo, PandaX:2024qfu, NEWS-G:2024jms, LZ:2024zvo}, green by XQC~\cite{McCammon:2002gb, Mahdawi:2018euy} and dotted gray by Lyman-$\alpha$ and Milky Way subhalos~\cite{DES:2020fxi, Nadler:2019zrb, Buen-Abad:2021mvc}.}
\end{figure*}

Figure~\ref{fig:heavy_mediator} presents our results for DM annihilating through a heavy mediator, considering both spin-dependent proton and spin-independent nucleon interactions. The top panels show the constraints derived from anomalous UV airglow, while the bottom panels display those from anomalous internal heating in Saturn, Uranus, Neptune, and Earth. We also project UV airglow sensitivity for a benchmark nearby Super-Jupiter and for anomalous heating at 0.1~kpc from the Galactic Center. New sensitivities for spin-dependent neutron scattering, as well as for DM subfractions, are shown in Appendix~\ref{sec:extra_xsec}.

For UV airglow in Fig.~\ref{fig:heavy_mediator}, our gas giant constraints on both spin-dependent and spin-independent scattering  surpass those from direct detection at low DM masses, benefiting from kinematic matching with hydrogen-rich planets like Jupiter and Saturn. This new sensitivity is possible even for this heavy mediator scenario, due to the part of the DM radial density profile already sitting in equilibrium in the atmosphere, where it is directly annihilating. The limits are bounded at low DM mass by the evaporation cutoff, and at high DM mass by the steep fall-off of $\fatm^\text{heavy}$ ($Cf.$ Fig.~\ref{fig:fatm_fint}).

For internal heat flow in Fig.~\ref{fig:heavy_mediator}, both Uranus and Neptune rule out some new parameter space, while those for the Earth and Saturn overlap with direct detection exclusions, providing an independent cross-check. For the Super-Jupiter, substantial new parameter space may be tested using DM heating once measurements with $e.g.$ JWST, Roman, or Rubin are made. The low-DM mass reach is again limited by evaporation, but is in addition limited due to lighter DM sitting increasingly in the atmosphere rather than the interior ($Cf.$ Fig.~\ref{fig:fatm_fint}). At higher masses, anomalous heating constraints are most sensitive near the typical planetary element masses, around 20~GeV for Earth. Beyond this, limits weaken roughly linearly with DM mass, similarly to direct detection. 

In Fig.~\ref{fig:heavy_mediator}, we include complementary constraints from direct detection experiments, which test the DM-SM scattering rate~\cite{CiaranDirectDetection}. Leading spin-dependent proton constraints are set by NEWS-G~\cite{NEWS-G:2024jms}, CRESST~\cite{CRESST:2022dtl}, LZ~\cite{LZ:2024zvo} and PICO-60~\cite{PICO:2019vsc}; and spin-independent nucleon interactions are constrained by CRESST~\cite{CRESST:2019jnq}, DarkSide-50~\cite{DarkSide:2022dhx}, PandaX-4T~\cite{PandaX:2024qfu}, and LZ~\cite{LZ:2024zvo}. Note that these experiments do not require an annihilation rate, unlike our results. At large cross sections, DM is expected to lose energy in the atmosphere or rock overburden before reaching deep underground detectors, producing a saturation limit (``ceiling'') for many direct detection experiments. Where available, we use the published upper limits; otherwise, we adopt ceiling values from Refs.~\cite{Buen-Abad:2021mvc, Bramante:2022pmn}. We also incorporate constraints from the XQC rocket experiment~\cite{McCammon:2002gb, Mahdawi:2018euy}. For a range of other constraints on benchmark models, see Refs.~\cite{Knapen:2017xzo,Cox:2024rew,Gori:2025jzu}. We omit Migdal-based constraints, because the Migdal effect has not been experimentally calibrated, and it remains unobserved using Standard Model probes~\cite{Xu:2023wev}.

Cosmological observations offer a complementary probe to direct detections experiments. In Fig.~\ref{fig:heavy_mediator}, we show limits from the Milky Way satellites; Lyman-$\alpha$ bounds are comparable to Milky Way satellite galaxy constraints~\cite{Rogers:2021byl} and are therefore not separately shown.

\section{\label{sec:light_mediator}Light Mediator Results}

We now investigate the case where DM annihilates into mediators that travel macroscopic distances before depositing energy into the surrounding medium. This scenario naturally arises when mediators are sufficiently light or long-lived. Mediators between the dark and visible sectors can acquire long lifetimes if their couplings to SM particles are suppressed. Examples of potential long-lived mediators include the dark photon and dark Higgs boson~\cite{FASER:2018eoc, Curtin:2018mvb}. Alternatively, mediators can inherit a large boost if the DM is comparably heavy, such that a long lifetime is not strictly required.
For simplicity, we refer to this scenario as the “light mediator” scenario.

Collider searches for missing energy, $e.g.$, at the LHC, can provide constraints on specific long-lived mediator models. Additionally, current and upcoming long-lived particle experiments such as FASER, MATHUSLA, SHiP, and CODEX-b aim to improve the sensitivity to particles with decay lengths exceeding $\mathcal{O}(100)$~m~\cite{Curtin:2018mvb}, which are otherwise difficult to detect at the LHC, by placing a detector some distance away from the beam collision point.

Planetary probes offer a complementary approach to test long-lived or light mediator properties. The observable fractions of DM energy deposited in the atmosphere and the planet interior, denoted as $\fatm^\text{light}$ and $\fint^\text{light}$, depend on both the mediator mass and its decay length $L$. The ability of different celestial bodies to probe various regions of parameter space arises from: (1) their size relative to $L$, (2) their characteristic DM radial distribution (see Fig.~\ref{fig:DMprofile}), and (3) their capture and annihilation rates. This interplay between planetary composition and mediator properties underscores the potential of planetary spectroscopy as a sensitive probe of physics beyond the SM.

\subsection{Energy Deposition via a Light Mediator}

We parameterize the properties of the mediator through the decay length
\begin{align}
    L \equiv \gamma \tau \ ,
\end{align}
where the Lorentz boost factor is approximately given by $\gamma \simeq m_\chi / m_\phi$, assuming the annihilating DM particles have a small relative velocity. We impose two main constraints on the mediator properties. First, its mass $m_\phi$ must lie between the electron mass and the DM mass to allow for on-shell production of the mediator and, subsequently, of visible final states. Second, its lifetime $\tau$ must be larger than $1/m_\phi$, though mediators with lifetimes larger than about a second could disrupt Big Bang Nucleosynthesis (BBN) by injecting energy into the primordial plasma~\cite{Chen:2009ab}.

Signals from DM annihilation into light mediators require the mediator to decay within the atmosphere (for airglow) or interior (for heating). We compute the resulting energy deposition using a Monte Carlo method that accounts for the DM radial distribution.

\subsubsection{Deposition in the Atmosphere}

\begin{figure}[t]
    \centering
    \includegraphics[width=0.9\columnwidth]{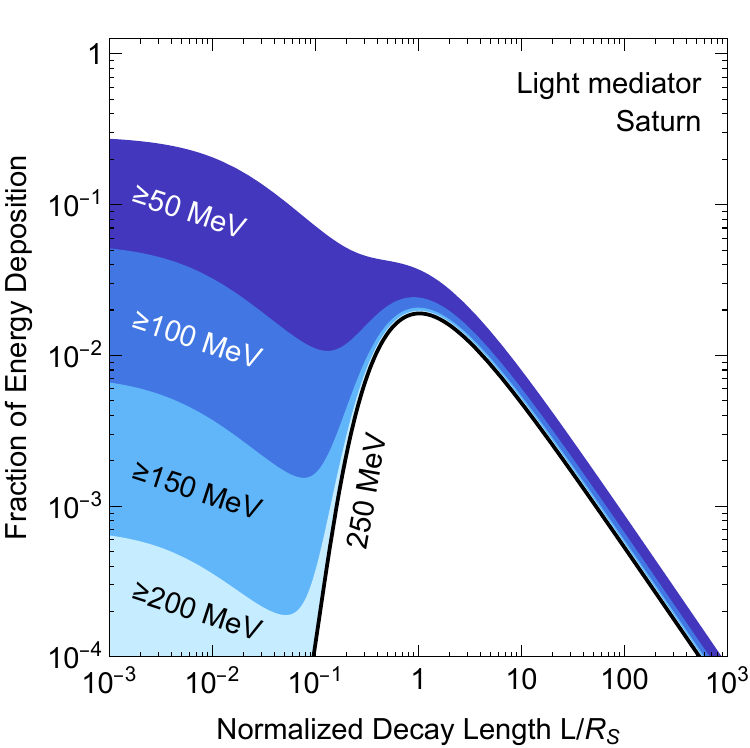}
    \caption{Fraction of energy deposited in Saturn's atmosphere from DM annihilation as a function of the boosted mediator decay length, normalized to Saturn's radius. The DM is radially distributed as in the LTE regime. DM masses above 250~MeV fall on the same low-lying line shown in thick black.}
    \label{fig:Saturn_fatm}
\end{figure}

\begin{figure*}[t]
    \centering
    \includegraphics[width=0.9\columnwidth]{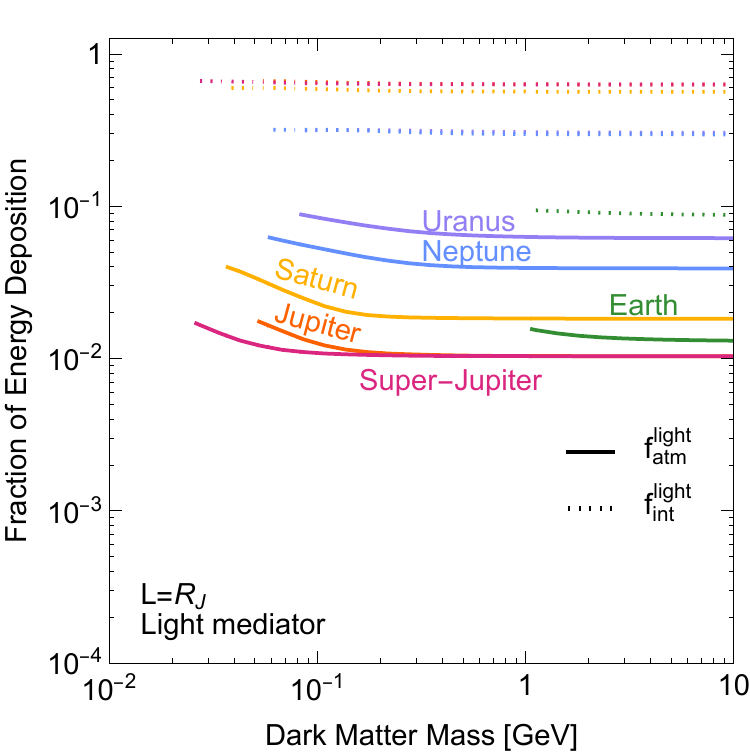}\hspace{4mm}
    \includegraphics[width=0.9\columnwidth]{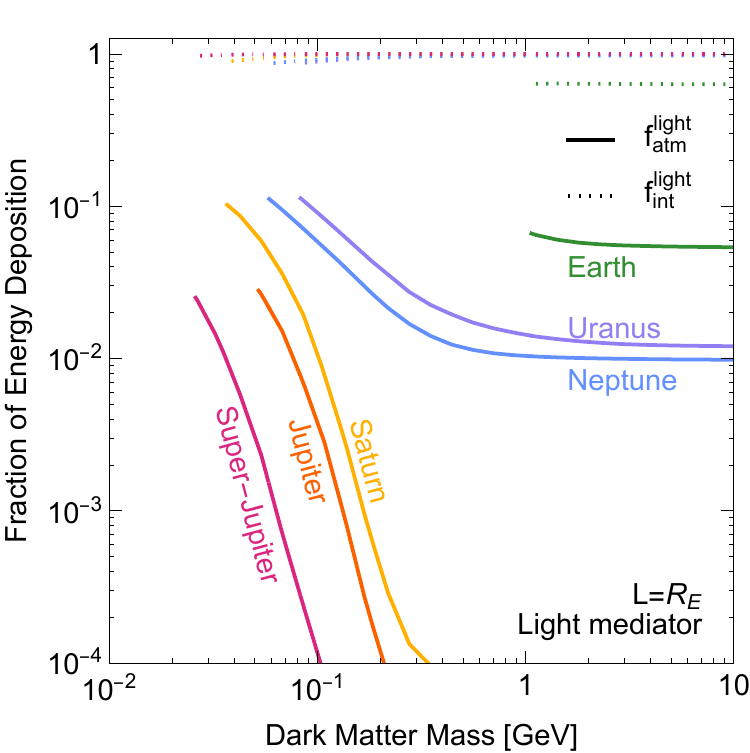}
    \caption{Fraction of energy deposited in the atmosphere ($\fatm^\text{light}$, solid) and interior ($\fint^\text{light}$, dotted) for DM annihilating through a light mediator, with DM distributed throughout the planet in the interior LTE regime, in the case of Jupiter, Saturn, Uranus, Neptune, Earth, and a Super-Jupiter. We present example decay lengths of Jupiter's radius, ${L = R_J}$~(left), and Earth's radius, ${L = R_E}$~(right), to illustrate the behavior with the mediator decay length and the complementarity of these planets.}
    \label{fig:fatm_light}
\end{figure*}

The atmospheric energy deposition fraction from DM annihilation into light mediators is computed as
\begin{align}\label{eq:fatm_light}
f_\text{atm}^\text{light} = \sum_i \mathbb{P}^\text{atm}_i / N \ ,
\end{align}
where $\mathbb{P}^\text{atm}_i$ is the probability that a mediator decays in the atmosphere, given an annihilation at point $i$, and $N$ is the total number of sampled points. Details of the Monte Carlo method are provided in Appendix~\ref{sec:mc-detail}.

In the limit of long decay lengths, a commonly used approximation is
\begin{align}\label{eq:fatm_approx}
\fatm^\text{light, approx} \simeq e^{-R/L} - e^{-R_\text{atm} / L} \ ,
\end{align}
assuming DM is concentrated at the planet's center. We find this approximation agrees well with the full result in Eq.~\eqref{eq:fatm_light} for DM masses $m_\chi \gtrsim 1$~GeV for Jupiter and $m_\chi \gtrsim 10$~GeV for Earth. For lighter masses, the DM core-concentration assumption breaks down, and the full calculation is required.

Figure~\ref{fig:Saturn_fatm} shows, in the example case of Saturn, how $\fatm^\text{light}$ depends on the mediator decay length in the large cross section regime (where the interior DM is in an LTE regime), for several DM masses; other planets show qualitatively similar behavior. In the small cross section regime, qualitatively similar results are obtained. At short decay lengths, energy deposition is largest for light DM, due to (1) the greater fraction of DM residing in the atmosphere, and (2) mediators decaying close to the production site, thereby tracing the DM density profile more closely. A local maximum appears near $L \sim R$, where mediators produced in the DM dense core are likely to decay in the atmosphere. For $L \gtrsim 10 R$, $\fatm^\text{light}$ falls off as $\sim 1/L$. As the DM mass increases, the atmospheric DM fraction decreases, suppressing $\fatm^\text{light}$ and bringing it closer to the approximation in Eq.~\eqref{eq:fatm_approx}.

\subsubsection{Deposition in the Interior}

The total fraction of energy deposited in the planetary interior is given by
\begin{align}\label{eq:fint_light}
f_\text{int}^\text{light} = \sum_i \mathbb{P}^\text{int}_i / N \ ,
\end{align}
where $\mathbb{P}^\text{int}_i$ is the probability that a mediator decays in the interior following an annihilation at point $i$.

An analytic approximation, similar to Eq.~\eqref{eq:fatm_approx}, is
\begin{align}\label{eq:fint_approx}
\fint^\text{light, approx} \simeq 1 - e^{-R/L} \ ,
\end{align}
which assumes that all DM is localized at the planetary center. Since interior deposition typically dominates over atmospheric deposition, this expression provides a good approximation across all DM masses considered, though our results use the full calculation.

\begin{figure*}
    \centering
    \includegraphics[width=0.32\textwidth]{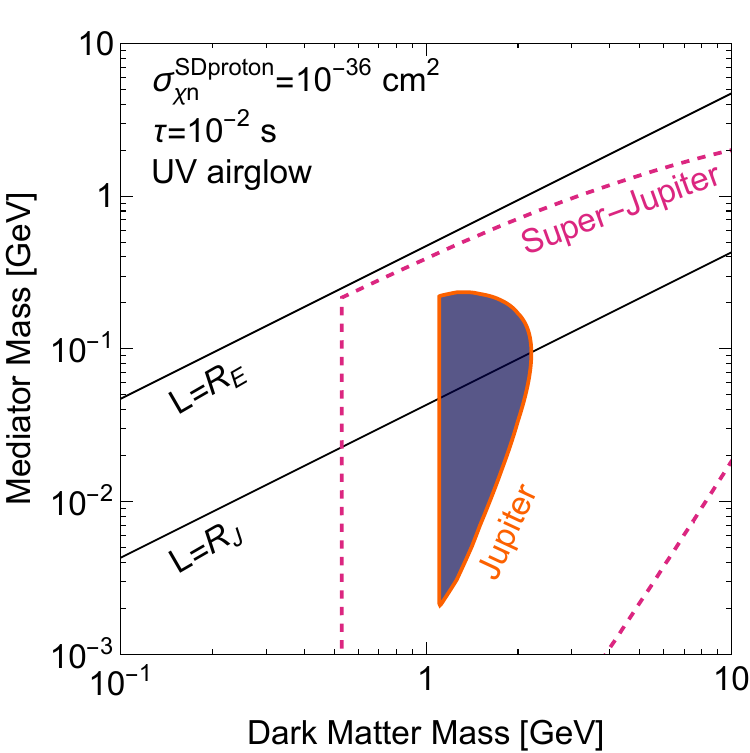}
    \includegraphics[width=0.32\textwidth]{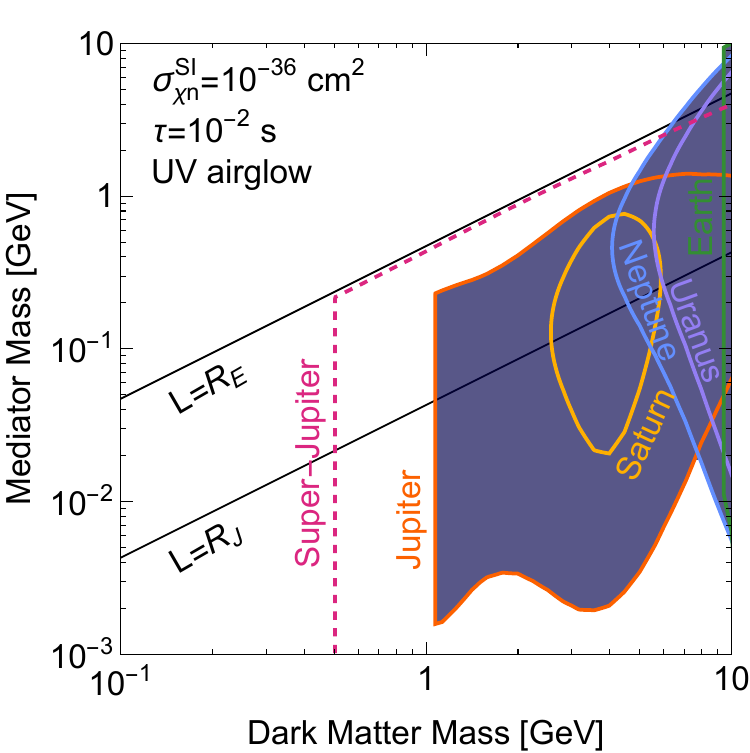}
    \includegraphics[width=0.32\textwidth]{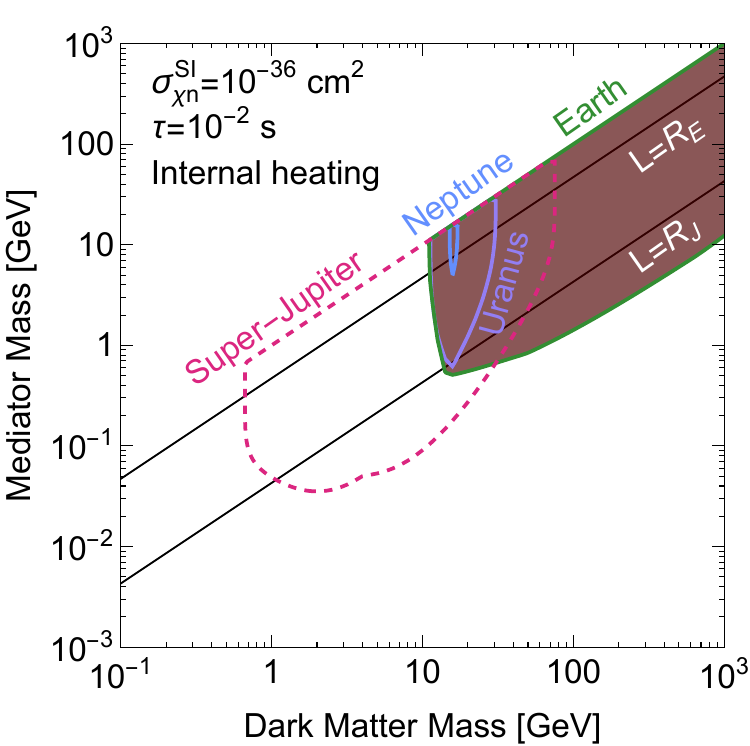}
    \caption{\label{fig:mphi_vs_mchi_m2}Parameter space constrained by planetary UV airglow and internal heat flow for a benchmark mediator lifetime of $\tau = 10^{-2}$~s and cross sections as labeled. The lines corresponding to the choices of $L$ from the panels of Fig.~\ref{fig:fatm_light} are shown in black.}
\end{figure*}

Figure~\ref{fig:fatm_light} shows the fraction of energy deposited in the atmosphere $\fatm^\text{light}$ and in the interior $\fint^\text{light}$ for a mediator with decay length of Jupiter's radius ${L = R_J}$~(left) and Earth's radius ${L = R_E}$~(right). As the right panel has a decay length which matches the Earth's radius, the Earth has the largest $\fatm^\text{light}$ when DM is located close to the core. Uranus and Neptune follow, as they have the next largest radii, followed by the gas giants. In the left panel, the Earth is then not the ``best'' candidate for the choice of decay length of Jupiter's radius. Note however that Jupiter still does not have the highest atmospheric energy deposition in the left panel, despite a decay length matched to its radius; this is because atmospheric thickness is also relevant, and other planets have larger atmospheric regions that can contribute to the signal.

\subsection{\label{sec:light_xsection}Airglow and Internal Heating Cross Section Constraints}

\begin{figure*}[t]
    \centering
    \includegraphics[width=0.44\textwidth]{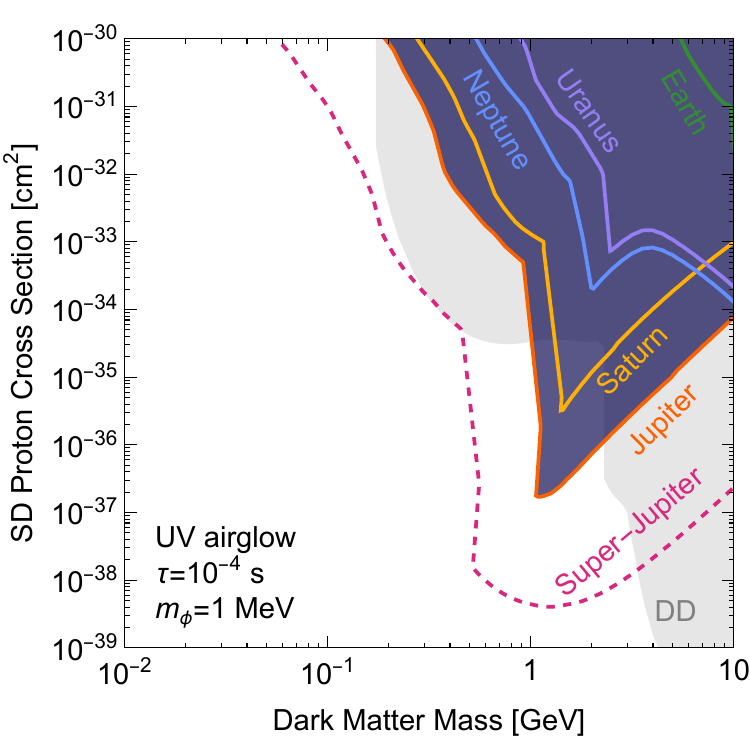}\hspace{4mm}
    \includegraphics[width=0.44\textwidth]{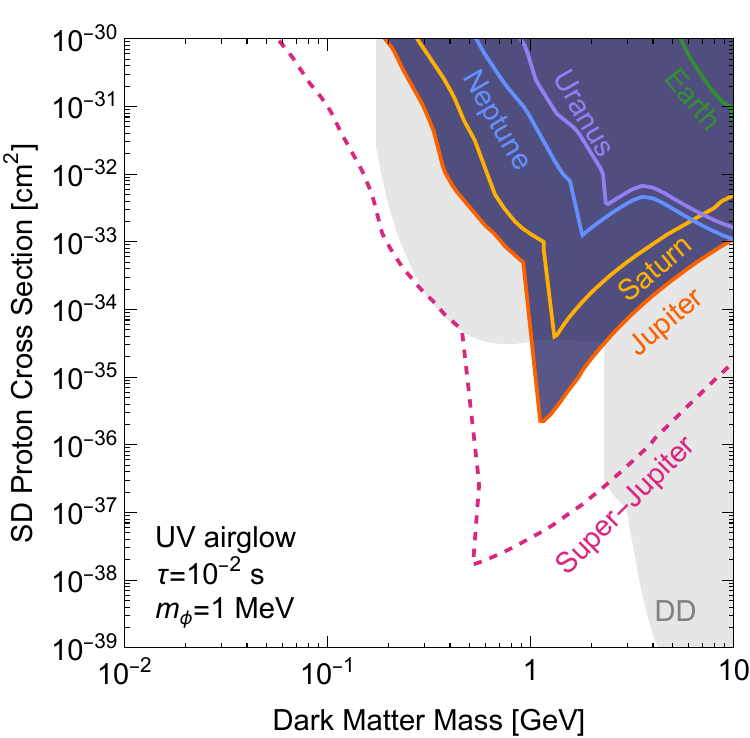}
    \caption{\label{fig:light_mediator_SDp_SI}Spin-dependent proton cross section constraints (solid, shaded) and future projections (dashed) using UV airglow for a benchmark 1~MeV light mediator with decay lifetime of $10^{-4}$~s (left) and $10^{-2}$~s (right). Complementary direct detection (DD) limits are shown in gray~\cite{PICO:2019vsc, CRESST:2019jnq, CRESST:2022dtl, DarkSide:2022dhx, LZ:2024zvo, PandaX:2024qfu, NEWS-G:2024jms, LZ:2024zvo}.}
\end{figure*}

Figure~\ref{fig:mphi_vs_mchi_m2} explores the mediator and DM mass parameter space for a spin-dependent proton and spin-independent interaction of $\sigma = 10^{-36}$~cm$^2$ and mediator lifetime of $\tau = 10^{-2}$~s, an arbitrary benchmark, for energy deposited in the atmosphere and interior. This highlights the complementary of different planets to probe dark sector properties. These can be linearly rescaled for other lifetimes, provided decay is kinematically possible. For reference, we also mark the decay lengths illustrated in Fig.~\ref{fig:fatm_light}. {The limits are cut off in the upper left region by requiring that the mediators $\phi$ to be produced on shell, $m_\phi < m_\chi$. The colored regions are truncated at small DM mass due to each planet's evaporation mass. At large DM mass, if $L \lesssim R$, the limits quickly weaken, as shown on the right panel of Fig.~\ref{fig:fatm_light} for Jupiter, Saturn, and a Super-Jupiter. The limits extend to smaller mediator mass when the main planet elements match the DM mass.}

Figure~\ref{fig:light_mediator_SDp_SI} shows instead constraints and projections in the DM-proton scattering vs DM mass plane, for a benchmark mediator mass of 1 MeV, and two benchmark mediator lifetimes of $10^{-2}$ and $10^{-4}$ seconds. We show here only UV airglow constraints, as the internal heating constraints are severely limited in the light mediator case. We find that sensitivities are strongest for mediator parameters when the energy deposition is largest in the relevant region of the planet, as expected. For results with additional benchmark points, and other cross section types, see Appendix~\ref{sec:extra_xsec}.

\section{\label{sec:summary}Summary and Conclusions}

We develop a comprehensive framework to probe DM interactions using planetary energy-injection signatures, focusing on UV airglow and internal heat flow. We consider two broad classes of DM annihilation scenarios: heavy mediators, which lead to prompt energy deposition near the annihilation site, and light or long-lived mediators, which result in displaced energy injection.

Within this framework, we derive general, model-independent expressions for the fraction of annihilation energy deposited into planetary atmospheres and interiors. We demonstrate that this fraction depends sensitively on the DM radial distribution, the planetary internal structure, and the mediator’s properties. Planetary size and composition also play key roles, especially across different mediator decay lengths and DM masses.

Using existing UV and heat flow measurements from Voyager, Cassini, New Horizons, AMS-02, CubeSats, and ground-based observations, we place new constraints on both spin-dependent and spin-independent DM–nucleon scattering cross sections. We also project sensitivities for a hypothetical free-floating Super-Jupiter near the Galactic Center, illustrating the promise of future planetary observations. These constraints are particularly powerful for DM masses below the reach of conventional direct detection experiments. Planetary energy injection signatures can also constrain cases of strongly-coupled DM subcomponents that would otherwise not reach underground detectors.

The most promising results are achieved using UV airglow, which enables sensitivity to low-mass DM well below traditional detection thresholds. Remarkably, even in the heavy mediator case, light DM can be well-probed due to the presence of thermalized DM in planetary atmospheres that can locally annihilate. In such scenarios, we find that more than $10\%$ of the captured DM can still annihilate in the atmosphere, greatly enhancing signal strength. For light mediators, the planetary environment provides unique access to macroscopic decay lengths inaccessible to terrestrial experiments, offering complementary reach to collider-based long-lived particle searches. In our companion UV airglow study, we assumed for simplicity that $100\%$ of captured DM annihilates in each planet's atmosphere{, corresponding to an optimistic, best-case scenario that maximizes the expected signal~\cite{Blanco:2024lqw}}. That analysis suggested Jupiter is generically the most sensitive planet. However, the more complete treatment presented here reveals that {the atmospheric annihilation fraction can vary significantly with the DM model and parameters, and reveals that Jupiter is not always the most sensitive target}.

Looking ahead, improvements in UV and infrared spectroscopy, both for Solar System planets and free-floating exoplanets, as well as more precise measurements of internal heat flows, could significantly extend the reach of this program; {see the companion paper~\cite{Blanco:2024lqw} for additional details on potential telescope and detection prospects}. Incorporating DM-electron scattering and additional energy-deposition signatures would further broaden sensitivity to a wider range of dark sector models.

{In this study, we adopt characteristic values of atmospheric parameters for the solar system planets, and their response to ionizing radiation. For example, we adopt a value of $C=\mathcal{O}(10)\,\text{R}~\mu\text{W}^{-1}\text{m}^{-2}$ as a characteristic value for the probable ionization of the gas-giant atmosphere by all the dark matter annihilation products, e.g., charged hadrons, heavy and light leptons. We also take benchmark locations for the atmospheric boundaries, between one bar of pressure and the exobase. Furthermore, we consider quasi-minimally ionizing electrons as a benchmark channel for ionization. However, it is important to note that the precise value of $C$ varies with energy, as well as the structure of the atmosphere and its constituents. A dedicated search for dark matter using this probe should simulate the development of the electromagnetic cascade, accounting for the salient features of the atmosphere, as well as the energy profile of the annihilation products.}

Overall, our results establish planetary energy-injection signatures as a powerful and versatile tool in the DM search program.  These signatures probe regions of parameter space unreachable by traditional methods, offering unique access to multiple model classes including light DM, strongly interacting subcomponents, as well as long-lived mediators with macroscopic decay lengths. This leverages the natural diversity of planetary environments to uncover the fundamental properties of dark matter, offering unique and complementary sensitivity compared to direct detection, collider searches, and cosmological probes.\\

\section*{Acknowledgments}

We thank Randy Gladstone, Wayne Pryor, Jacques Gustin, and Aaron Vincent for helpful discussions, and thank Ciaran O'Hare for his public repository of DM direct detection bounds that we utilized. C.B. is supported in part by NASA through the NASA Hubble Fellowship Program grant HST-HF2-51451.001-A awarded by the Space Telescope Science Institute, which is operated by the Association of Universities for Research in Astronomy, Inc., for NASA, under contract NAS5-26555. R.K.L. and J.T. are supported by the U.S. Department of Energy under Contract DE-AC02-76SF00515. M.M. is supported by subMIT at MIT Physics, the U.S. Department of Energy under contract DE-SC0012567, and the Simons Foundation (Grant No. 929255).

\appendix
\onecolumngrid

\section{\label{sec:properties}Planetary Properties, Composition, and Uncertainties}

Table~\ref{tab:planet_profiles} presents the SM density and temperature profiles for the interiors of the giant planets and Earth. We adopt the 1~bar pressure level to separate interior and atmosphere, denoting the 1~bar radius as $R$ and the exobase height as $h_\text{exo}$. The Super-Jupiter model uses Jupiter’s radius but assumes a tenfold increase in the density profile (and thus total mass) and a fivefold increase in the temperature profile; these values can be adjusted once an observational target is identified. For all bodies, we assume isotropic density, temperature, and composition profiles as in our companion paper~\cite{Blanco:2024lqw}, and neglect the small planetary oblateness, treating them as perfect spheres.

Table~\ref{tab:planet_data} lists the properties and isotopic compositions used to compute DM capture. A dash ($-$) denotes negligible elemental abundance.

Table~\ref{tab:isotopes} summarizes the non-zero nucleus spin and proton/neutron spin expectation values adopted for spin-dependent interactions, taken from Refs.~\cite{Bednyakov:2004xq, Hooper:2018bfw, Bramante:2019fhi}. Elements not listed (${}^4$He, ${}^{12}$C, ${}^{16}$O) have $J=\ev*{S_{\lbrace p,n \rbrace}}=0$ and thus do not contribute.

\begin{table}[t]
    \caption{\label{tab:planet_profiles}Density and temperature profile references for the interiors of the giant planets and Earth.}
    \centering
    \begin{tabular}{p{2.2cm}
    p{3.6cm}}
        \toprule
        & Interior \\ \midrule 
        Jupiter 
        & model J11-4a from \cite{2012ApJS2025F} \\
        Saturn 
        & baseline model from \cite{2021NatAs...5.1103M} \\
        Uranus
        & \cite{Mitra:2004fh} \\
        Neptune
        & model N1 from \cite{Nettelmann:2012su}\\
        Earth 
        & \cite{Dziewonski:1981xy, Neufeld:2018slx} \\
        \bottomrule
    \end{tabular}
\end{table}

\begin{table}[H] 
    \centering    
    \caption{\label{tab:planet_data}List of parameters used for the giant planets and Earth: radius $R$, mass $M$, exobase height $h_{\rm exo}$, 
    and elemental abundance of the most common elements and isotopes in mass fraction used for capture (Sec.~\ref{sec:capture}) and evaporation (Sec.~\ref{sec:evaporation}). The fractional composition in any column might not add up to one due to rounding. In the last line, $m_\text{SM}$ is the SM target mass used in the LTE radial profile, Eq.~\eqref{eq:DM_radial_profile}. Exobase heights can be estimated for Jupiter~\cite{2001Icar..152..151A}, Saturn~\cite{2013Icar..226.1318K}, Uranus and Neptune~\cite{2020RSPTA.37890478M}, Earth~\cite{1960JGR....65.2577S}, and Jupiter-like planets~\cite{2004A&A...418L...1L}.} 
    \begin{tabular}{l >{\raggedleft}p{1.8cm} >{\raggedleft}p{1.8cm} >{\raggedleft}p{1.8cm} >{\raggedleft}p{1.8cm} r} 
    \toprule 
    & Jupiter & Saturn & Uranus & Neptune & Earth \\
    \midrule 
    $R$ ($\times 10^7$~m) & 6.99 & 5.83 & 2.54 & 2.46 & 0.6371 \\ 
    $M$ ($\times 10^{26}$~kg) & 18.98 & 5.68 & 0.86 & 1.02 & 0.0597 \\
    $h_\text{exo} \, (\times 10^3~ \text{km})$& $\sim$2& 3.0& 6.5& 4&1.0\\ 
    \midrule
    {[H]} & 0.75 & 0.75 & 0.18 & 0.18 & $-$ \\ 
    {[He]} & 0.25 & 0.25 & $-$ & $-$ & $-$ \\ 
    {[C]} & $-$ & $-$ & 0.25 & 0.25 & $-$ \\ 
    {[N]} & $-$ & $-$ & 0.27 & 0.27 & $-$ \\ 
    {[O]} & $-$ & $-$ & 0.30 & 0.30 & 0.30 \\ 
    {[Mg]} & $-$ & $-$ & $-$ & $-$ & $0.15$ \\
    {[Al]} & $-$ & $-$ & $-$ & $-$ & $0.015$ \\
    {[Si]} & $-$ & $-$ & $-$ & $-$ & $0.14$ \\
    {[Ca]} & $-$ & $-$ & $-$ & $-$ & $0.017$ \\
    {[Fe]} & $-$ & $-$ & $-$ & $-$ & $0.32$ \\
    {[${}^2$H]} & $7.5\times 10^{-6}$ & $7.5\times 10^{-6}$ & $1.8\times 10^{-6}$ & $1.8\times 10^{-6}$ & $-$ \\ 
    {[${}^3$He]} & $2.5\times 10^{-5}$ & $2.5\times 10^{-5}$ & $-$ & $-$ & $-$ \\ {[${}^{13}$C]} & $-$ & $-$ & $2.5\times 10^{-5}$ & $2.5\times 10^{-5}$ & $-$ \\ 
    {[${}^{15}$N]} & $-$ & $-$ & $1.1\times 10^{-3}$ & $1.1\times 10^{-3}$ & $-$ \\ 
    {[${}^{17}$O]} & $-$ & $-$ & $1.2\times 10^{-4}$ & $1.2\times 10^{-4}$ & \phantom{A}$1.2\times 10^{-4}$ \\ 
    {[${}^{25}$Mg]} & $-$ & $-$ & $-$ & $-$ & $0.015$ \\ 
    {[${}^{29}$Si]} & $-$ & $-$ & $-$ & $-$ & $0.0066$ \\ 
    {[${}^{43}$Ca]} & $-$ & $-$ & $-$ & $-$ & $2\times 10^{-5}$ \\ 
    {[${}^{57}$Fe]} & $-$ & $-$ & $-$ & $-$ & $0.007$ \\ 
    \midrule
    $m_\text{SM}$ ($\times m_p$) & 1 & 1 & 12 & 12 & 16 \\
    \bottomrule 
    \end{tabular} 
\end{table}

\begin{table}[H] 
    \centering 
    \caption{\label{tab:isotopes}Nucleus spin {$J$} and expectation values {$\ev{S_p}$ and $\ev{S_n}$} of the proton and neutron spins, for elements relevant for spin-dependent scattering.}
    \begin{tabular}{c c c c c c c c c c c c c}
    \toprule 
    & \phantom{A}H\phantom{A} & \phantom{A}${}^2$H\phantom{A} & \phantom{A}${}^3$He\phantom{A} & \phantom{A}${}^{13}$C\phantom{A} & \phantom{A}${}^{14}$N\phantom{A} & \phantom{A}${}^{15}$N\phantom{A} & \phantom{A}${}^{17}$O\phantom{A} & \phantom{A}${}^{25}$Mg\phantom{A} & \phantom{A}${}^{27}$Al\phantom{A} & \phantom{A}${}^{29}$Si\phantom{A} & \phantom{A}${}^{43}$Ca\phantom{A} & \phantom{A}${}^{57}$Fe\phantom{A} \\
    \midrule 
    $J$ & 1/2 & 1 & 1/2 & 1/2 & 1 & 1/2 & 5/2 & 5/2 & 5/2 & 1/2 & 7/2 & 1/2 \\ 
    \phantom{A}$\ev*{S_p}$\phantom{A} & 0.5 & 0.5 & $-0.081$ & 0 & 0.5 & $-0.136$ & $-0.008$ & 0.04 & 0.333 & 0.054 & 0 & 0 \\ 
    $\ev*{S_n}$ & 0 & 0.5 & 0.552 & $-0.183$ & 0.5 & 0.028 & 0.48 & 0.376 & 0.043 & 0.204 & 0.5 & 0.5 \\
    \bottomrule 
    \end{tabular} 
\end{table}

The outer planets’ atmospheres are reasonably well characterized by Voyager and later missions such as Galileo, Cassini, and Juno through infrared spectra and radio occultation~\cite{1981JGR....86.8713G, 1984ApJ...282..807C, 2022Icar..37814937H}. In Jupiter, the helium mass fraction is $\sim 0.25$, consistent with primordial values and indicating minimal helium differentiation between atmosphere and interior~\cite{1981JGR....86.8713G}. Saturn shows evidence of helium separation: Voyager data suggested $\sim0.18$--$0.25$~\cite{2000Icar..144..124C}, while Cassini found $\sim0.075$--$0.23$ depending on instrument~\cite{2020PSJ.....1...30A}. For consistency, we adopt primordial helium abundances for both Jupiter and Saturn, as summarized in Table~\ref{tab:planet_data}. For Uranus and Neptune, their compositions are less certain. Their outer envelopes contain helium, but the interior composition is uncertain and may include mixed ices (H$_2$O, NH$_3$, CH$_4$), a single dominant ice, or even rocky material such as SiO$_2$~\cite{1987JGR....9215003C, 1987RvGeo..25..251B, 2018P&SS..155...12M}. Here we adopt a mixed-ice interior model, detailed in Table~\ref{tab:planet_data}.
 
While there are uncertainties in planetary compositions and interiors, our strongest sensitivities generally come from Super-Jupiters, Jupiter, and Saturn, which are hydrogen-rich. Even deliberately conservative assumptions---e.g., reducing the hydrogen fraction to $75\%$ and removing helium and heavier elements---change our SI and SD proton-scattering bounds by only tens of percent, demonstrating that they are largely insensitive to plausible composition variations. For Uranus and Neptune, similarly, the main elements composing it are known to an order one factor. The main exception is SD neutron scattering in the gas giants, which can depend on trace isotopes such as $^2$H and $^3$He and is therefore more sensitive to composition systematics. Another source of uncertainty is the core temperature: varying it shifts the evaporation cutoff approximately linearly in DM mass. This effect is straightforward to rescale and is expected to be known within a factor of a few for Super-Jupiters and better for Solar System planets.

Overall, by comparing with very conservative assumptions, such as considering only minimal fractions of the main elements or excluding heavier elements entirely, we find that our strongest SI and SD proton-scattering constraints remain fairly robust against uncertainties in planetary composition. The main caveat is for SD neutron scattering, where sensitivity depends on low-abundance isotopes and is thus more composition dependent.


\section{\label{sec:wave}Results for Benchmark Annihilation Rates and Capture-Annihilation Equilibrium}

\begin{figure}[t]
    \centering
    \includegraphics[width=0.44\linewidth]{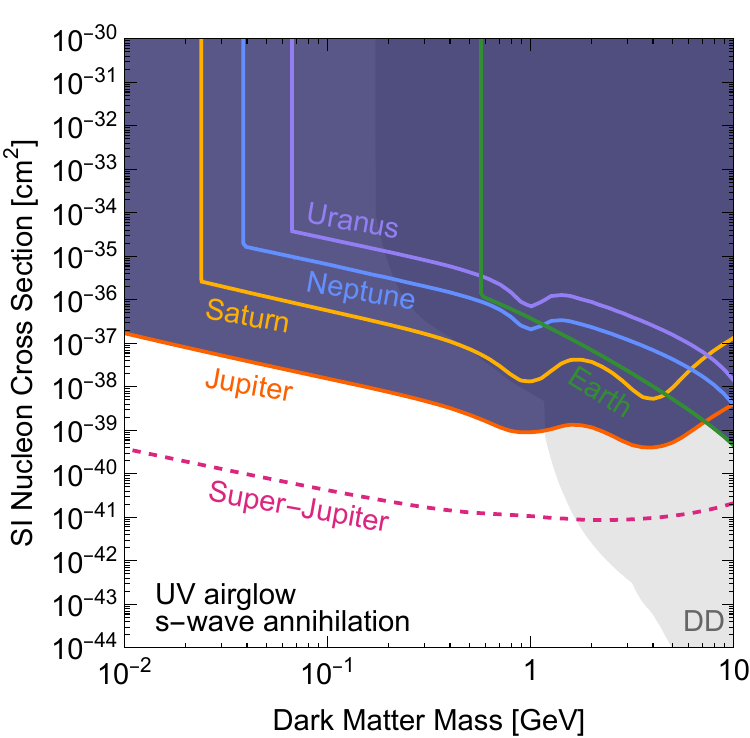}\hspace{4mm}
    \includegraphics[width=0.44\linewidth]{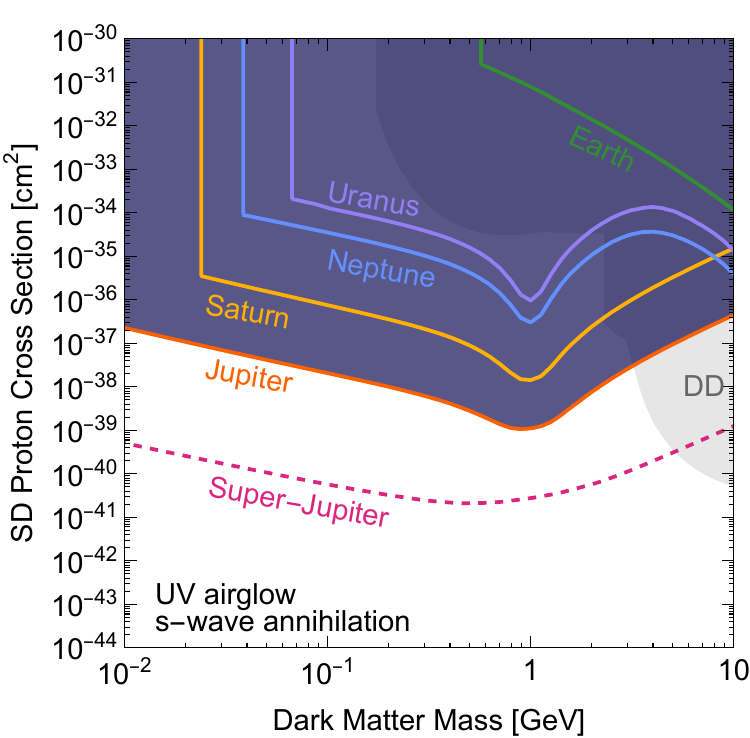}
        \includegraphics[width=0.44\linewidth]{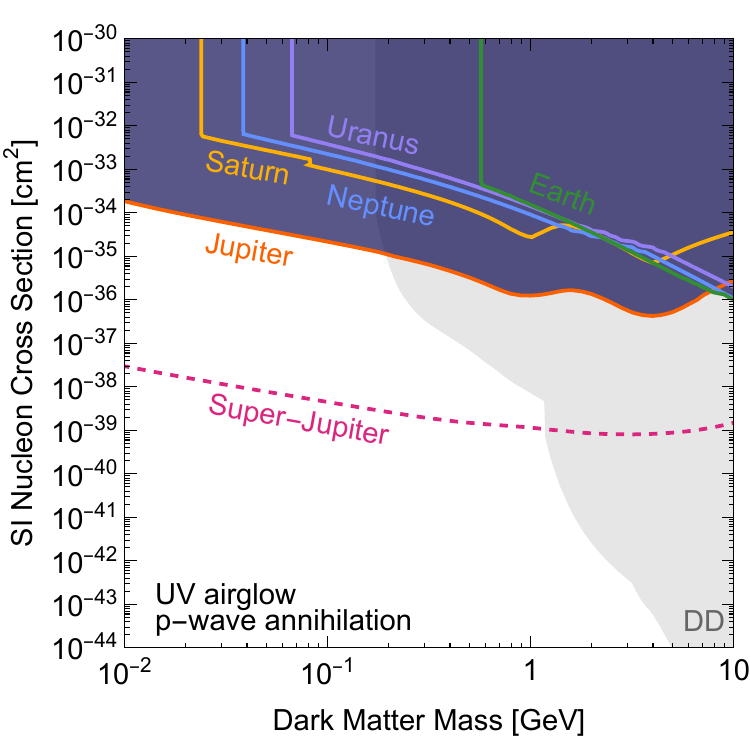}\hspace{4mm}
    \includegraphics[width=0.44\linewidth]{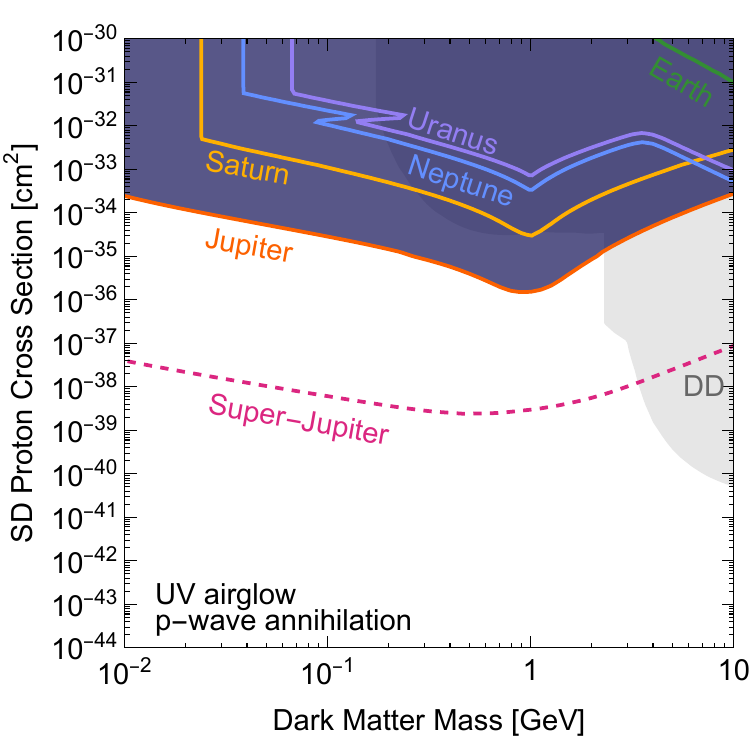}
    \caption{\label{fig:equilibrium} Optimal UV airglow limits including a factor accounting for potential non-equilibration with $s$-wave annihilation (top row), and $p-$wave annihilation (bottom row). The left panel is for a spin-independent interaction and the right panel, for a spin-dependent proton interaction. Grey represents the combined most recent direct detection~(DD) constraints~\cite{PICO:2019vsc, CRESST:2019jnq, CRESST:2022dtl, DarkSide:2022dhx, LZ:2024zvo, PandaX:2024qfu, NEWS-G:2024jms, LZ:2024zvo}.}
\end{figure}

Even in the scenario where DM capture-annihilation equilibrium is not reached, detectable signals are possible, due to a sufficiently large DM population accumulated inside the planet. We therefore now  investigate the parameter space for two benchmark scenarios of the thermally-averaged annihilation cross section: (1) $s$-wave rates of ${\ev{\sigma v}_s \simeq 3 \times 10^{-26}}$~cm$^3$/s, and as well as (2) $p$-wave annihilation rates, which are suppressed compared to $s-$wave rates by the DM thermal velocity squared, $v_\text{th}^2$.

To do this consistently, the equilibrium condition ${C_\text{cap} = N_\chi^2 C_\text{ann}}$ is replaced by ${C_\text{cap} \tanh^2(t / t_\text{eq}) = N_\chi^2 C_\text{ann}}$, which allows for a calculation of the rates both in and out of capture-annihilation equilibrium. Here we take the equilibration timescale $t_\text{eq}$ as per Eq.~\eqref{eq:t_eq}, and take time $t$ to be 4.5~Gyr in the Solar System planets and 10~Gyr for the Super-Jupiter.
For $p$-wave annihilation, we define the DM temperature as the volume-averaged temperature within the volume bounded by the scale radius $r_\chi$, defined in Eq.~\eqref{eq:rchi}. We define the annihilation rate as $C_\text{ann} = \ev{\sigma v} / ((2\pi)^{3/2} r_\chi^3)$, where the denominator is the volume enclosed within the thermal radius. The scale radius $r_\chi$ is defined in the main text, while the thermal radius is $r_\text{th} = \sqrt{9 T_c / (4 \pi G \rho_c m)}$~\cite{Sanematsu:2015kvk, Acevedo:2020gro, Chu:2024gpe, Luo:2025psd}.

Figure~\ref{fig:equilibrium} presents our optimal (${\fatm \sim 1}$) UV airglow limits for the gas giants, ice giants, Earth, and a local Super-Jupiter for the $s$- and $p$-wave annihilation rates, but without the effect of evaporation. We see that $s$-wave sensitivity is possible with all objects. For $p$-wave annihilation, it is not reached everywhere for all objects, but the minimum sensitivities shown here show that some of the parameter space is also achievable. The vertical lines indicate where the scale radius $r_\chi$ reaches the radius of the planet, beyond which we do not define the isothermal radial profile nor the equilibration time. Complementary constraints from direct detection are shown, and are the same as those cited in the main text.

\begin{figure}[ttt]
    \centering
    \includegraphics[width=0.44\linewidth]{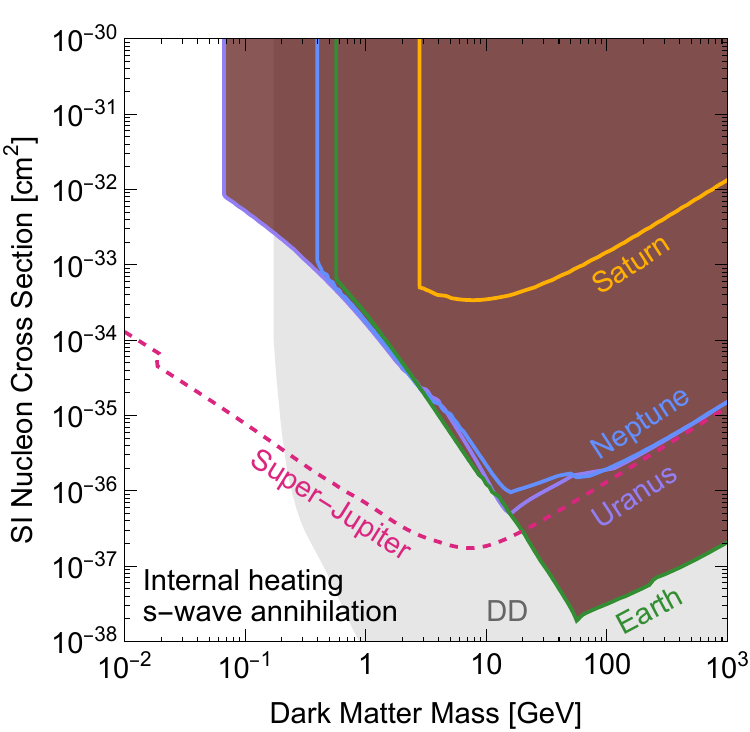}\hspace{4mm}
    \includegraphics[width=0.44\linewidth]{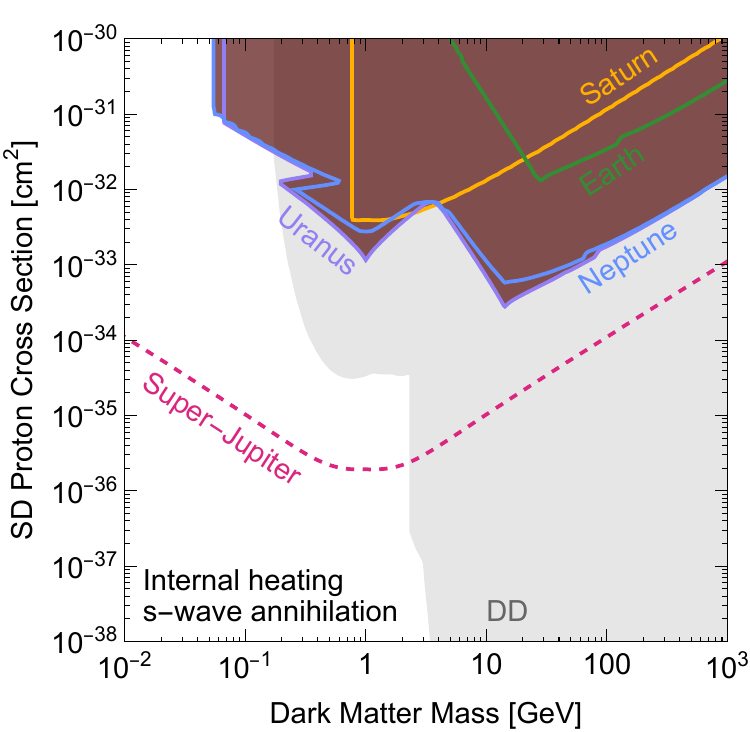}
        \includegraphics[width=0.44\linewidth]{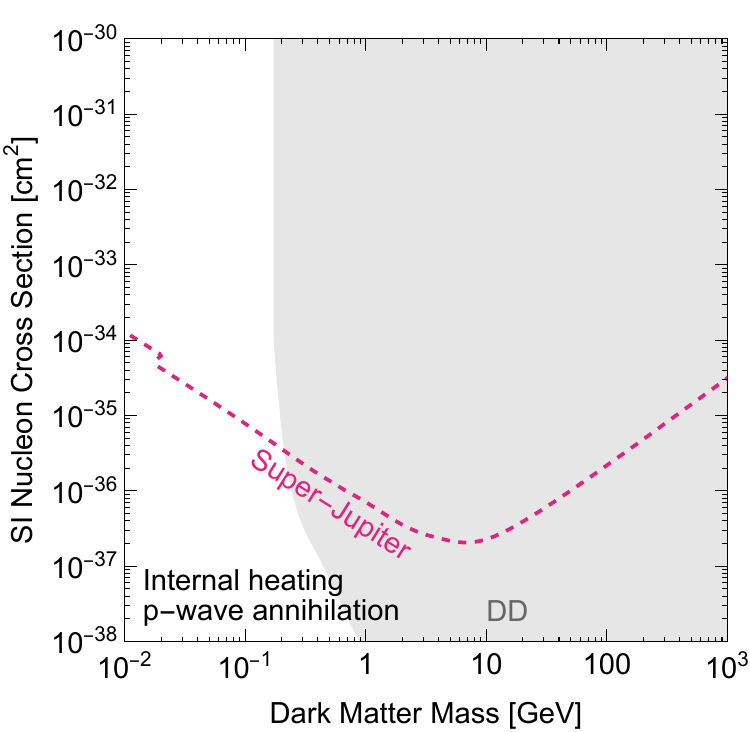}\hspace{4mm}
    \includegraphics[width=0.44\linewidth]{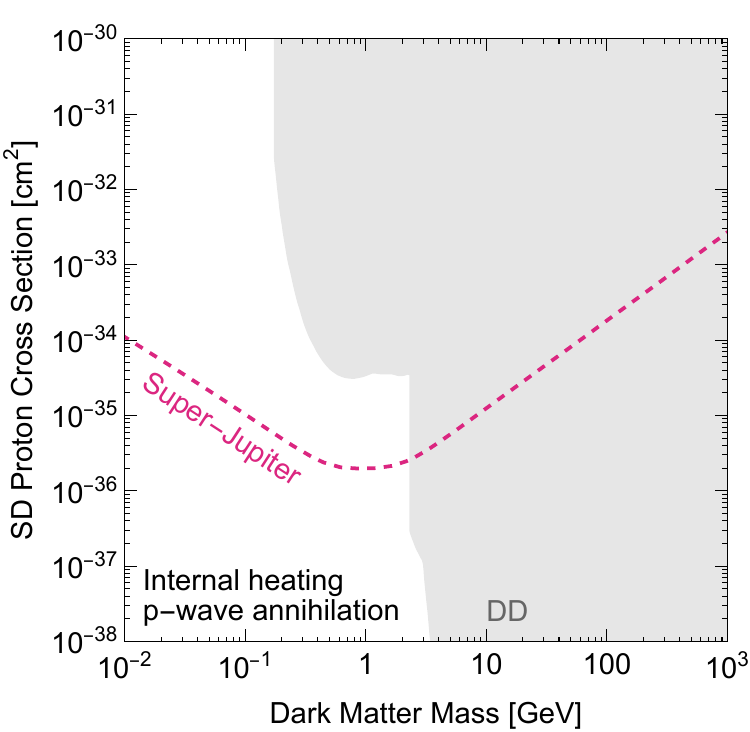}
    \caption{\label{fig:equilibrium_heating} Optimal internal heating limits including a factor accounting for potential non-equilibration with $s$-wave annihilation (top row), and $p-$wave annihilation (bottom row). The left panel is for a spin-independent interaction and the right panel, for a spin-dependent proton interaction. Grey represents the combined most recent direct detection~(DD) constraints~\cite{PICO:2019vsc, CRESST:2019jnq, CRESST:2022dtl, DarkSide:2022dhx, LZ:2024zvo, PandaX:2024qfu, NEWS-G:2024jms, LZ:2024zvo}.
    }
\end{figure}

Figure~\ref{fig:equilibrium_heating} shows the ${\fint \sim 1}$ internal heating limits for the five Solar System planets and a Super-Jupiter located in the Galactic center, assuming the standard $s$- and $p$-wave WIMP thermally-averaged relic annihilation cross section and without the effect of evaporation. We again see that $s$-wave sensitivity is possible with all objects, while $p$-wave sensitivity via internal heating is only achieved with the Super-Jupiter. Note that in Figs.~\ref{fig:equilibrium} and~\ref{fig:equilibrium_heating}, the discontinuity around $10^{-32}~$cm$^2$ for Uranus and Neptune is a numerical artifact of an approximation for switching between single and multi-scatter regimes with \texttt{Asteria}, and is not a physical feature; the physical result in that region will be within about a factor 2 of what is shown.  

Our results apply to $s$- and $p$-wave annihilation, though the $s$-wave annihilation cross section can be constrained by cosmological observations, under some assumptions. In particular, measurements of the cosmic microwave background~(CMB) by the Planck satellite can set stringent limits on DM annihilation, as DM-induced energy injection into the photon-baryon plasma can affect the ionization history and CMB anisotropies~\cite{Slatyer:2015jla, Leane:2018kjk, Planck:2018vyg}. Under the assumption of $2\rightarrow2$ annihilation fully into visible final states, DM with a mass below $\sim 10$~GeV is constrained~\cite{Leane:2018kjk}. However, these bounds can be weakened for a range of WIMP models, including non $2\rightarrow2$ processes such as those from hidden sector WIMPs~\cite{Abdullah:2014lla,Elor:2015bho,Bell:2016fqf,Bell:2016uhg,Bell:2017irk}, or a non-zero branching fraction into neutrinos. 

\section{Additional Cross Section Results}
\label{sec:extra_xsec}

\subsection{Heavy Mediator}

\subsubsection{\label{sec:spin-dependent}Spin-Dependent Neutron Interactions}

\begin{figure*}[t]
    \centering
    \includegraphics[width=0.45\textwidth]{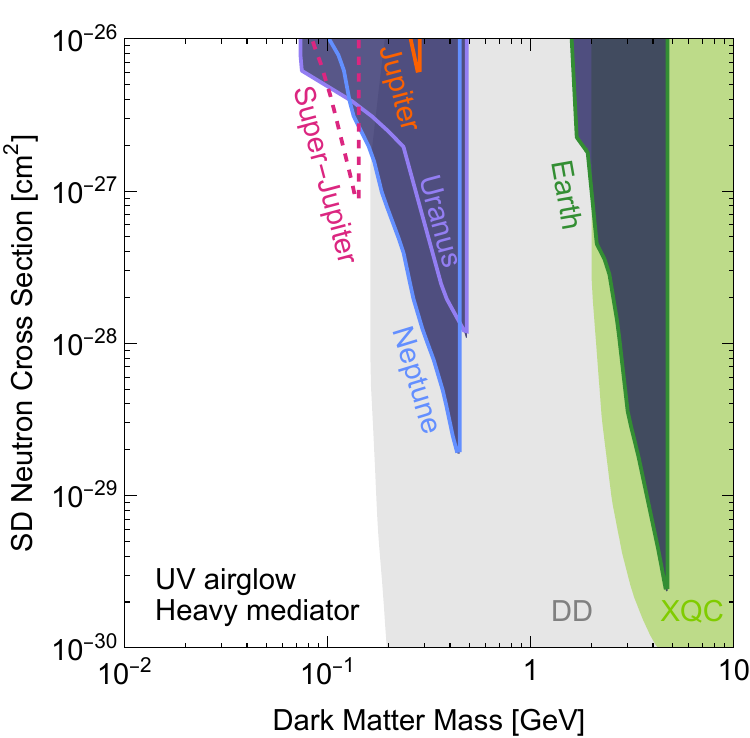}
    \includegraphics[width=0.45\textwidth]{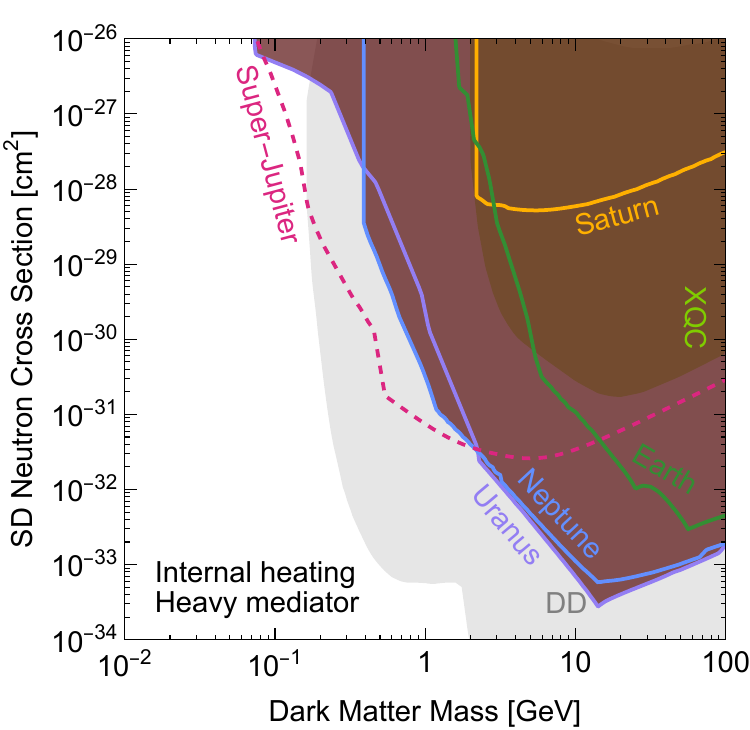}
    \caption{\label{fig:heavy_mediator_neutron}
     Constraints (solid, shaded) and future projections (dashed) on spin-dependent neutron scattering, assuming DM annihilation through a heavy mediator using planetary UV airglow (left) and internal heat flow (right). Gray zone is constrained by direct detection (DD) experiments~\cite{SuperCDMS:2018gro, CRESST:2022dtl, LZ:2024zvo, XENON:2025vwd}, green is XQC~\cite{McCammon:2002gb, Mahdawi:2018euy}.}
\end{figure*}

Figure~\ref{fig:heavy_mediator_neutron} presents heavy-mediator constraints on spin-dependent DM-neutron interactions, using the Solar System giant planets, Earth, and a Super-Jupiter, for both UV airglow and internal heating.
Direct detection experiments constrain these spin-dependent neutron-DM scattering interactions. The strongest constraints come from CRESST~\cite{CRESST:2022dtl}, CDMSlite~\cite{SuperCDMS:2018gro}, LZ~\cite{LZ:2024zvo} and {XENONnT}~\cite{XENON:2025vwd}. The CMB, BAO, and \mbox{Lyman-$\alpha$} forest cannot constrain a neutron-only spin-dependent interaction, leaving the parameter space unavailable to direct detection due to the experiments' shielding poorly probed.

\subsubsection{DM Subfraction}

The scenario where only a subfraction of particle DM interacts non-gravitationally with SM particles has been of increasing interest recently, $e.g.$~\cite{Majorana:2023ecz}, due to the prediction in some model classes, $e.g.$~\cite{Pospelov:2007mp, Cyr-Racine:2012tfp}, as well as the inability of existing experiments to easily probe this regime. In these scenarios, cosmological and astrophysical constraints become less stringent. Specifically, constraints from the CMB, BAO, \mbox{Lyman-$\alpha$} forest, and Milky Way satellite galaxy bounds can disappear entirely when ${f_\chi \lesssim 2 \times 10^{-3}}$~\cite{Boddy:2018wzy}, making it an interesting scenario for lab-based experiments.

To study this scenario, we define ${f_\chi < 1}$ to be the fraction of DM which interacts non-gravitationally with SM particles and we rescale Eq.~\eqref{eq:C_cap} by $C_\text{cap} \rightarrow f_\chi C_\text{cap}$.

\begin{figure}[t]
    \centering
    \includegraphics[width=0.45\linewidth]{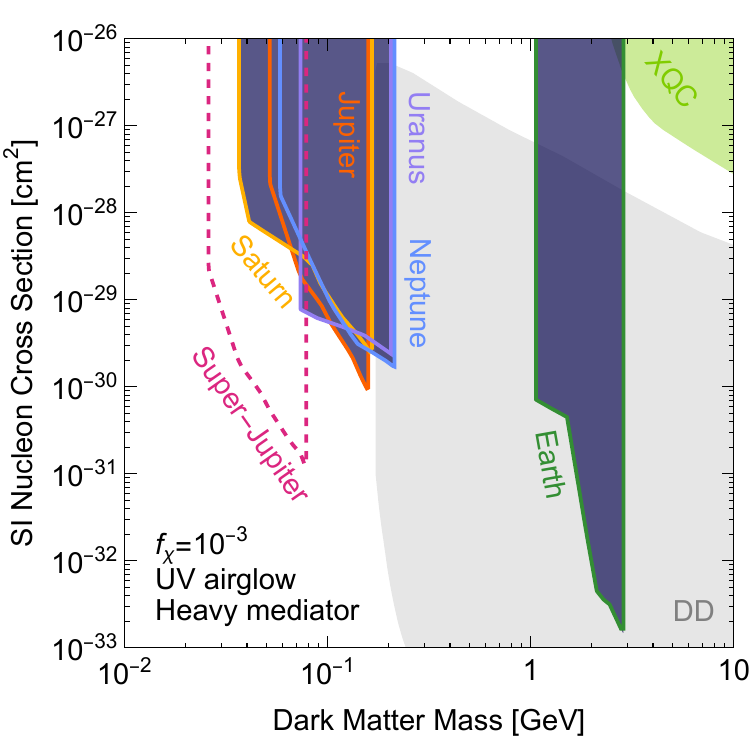}
    \includegraphics[width=0.45\linewidth]{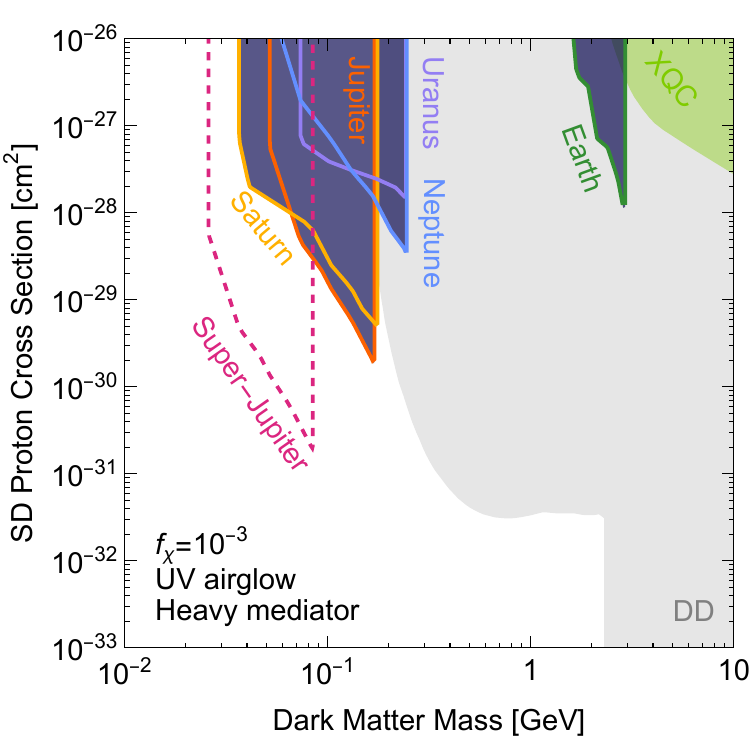}
    \caption{Constraints (solid, shaded) and future projections (dashed) on spin-independent nucleon scattering (left) and
spin-dependent proton scattering (right), assuming a DM subfraction ${f_\chi = 10^{-3}}$ annihilating through a heavy mediator, using planetary UV airglow. Gray zone is constrained by direct detection (DD) experiments~\cite{SuperCDMS:2018gro, CRESST:2022dtl, LZ:2024zvo, XENON:2025vwd}, green is XQC~\cite{McCammon:2002gb, Mahdawi:2018euy}.}
    \label{fig:subfraction}
\end{figure}

Figure~\ref{fig:subfraction} presents the UV airglow constraints that can be set on a subfraction $f_\chi = 10^{-3}$ of DM interaction through spin-independent scattering with nucleons. {While the capture rate scales linearly with $f_\chi$, the resulting constraints do not necessarily weaken by a factor $1/f_\chi$ across all parameter space. At low cross sections, the DM population is limited by evaporation rather than capture. Since the evaporation mass is independent of $f_\chi$, in this regime reducing $f_\chi$ leads to a much weaker rescaling of the constraints.}

The disappearing cosmology bounds (Lyman-$\alpha$ forest, Milky-Way satellite galaxies, CMB+BAO) open up parameter space at large cross sections which cannot be probed by underground DM experiments, though our bounds for Earth can be fully probed via DM-induced power in quantum devices~\cite{Das:2022srn,Das:2024jdz}; our limits with other planets can reach lower DM masses. While the XQC rocket experiment and lower limit of direct detection clearly move up by three orders of magnitude, the UV airglow constraints adopt a slower change {due to the evaporation-dominated regime}. For comparable subfractions, we find that no constraints are obtained using internal heating in Solar System planets. 

\subsection{Light Mediator}

In addition to the results of the main text, we present here light mediator constraints and sensitivities for additional parameters.

\begin{figure*}
    \centering
    \includegraphics[width=0.32\textwidth]{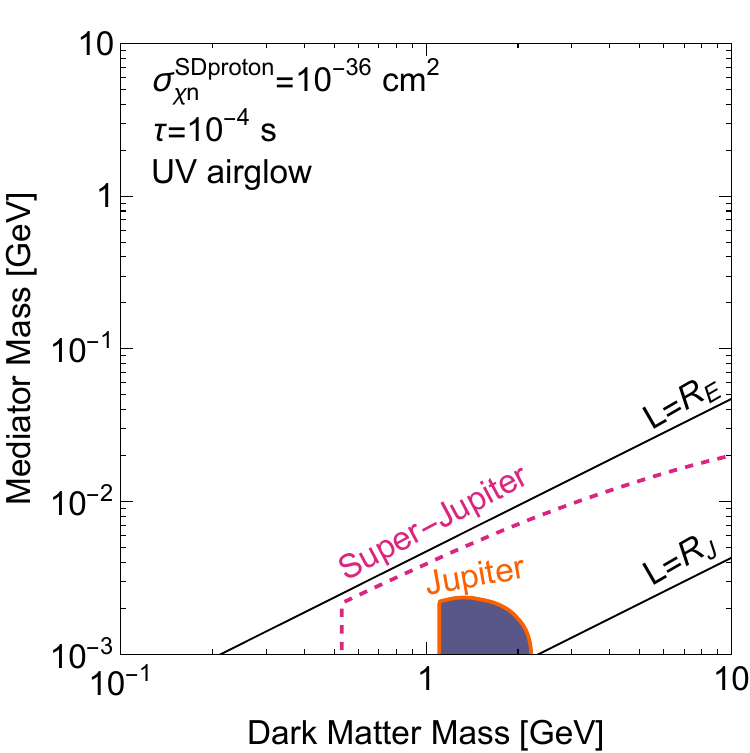}
    \includegraphics[width=0.32\textwidth]{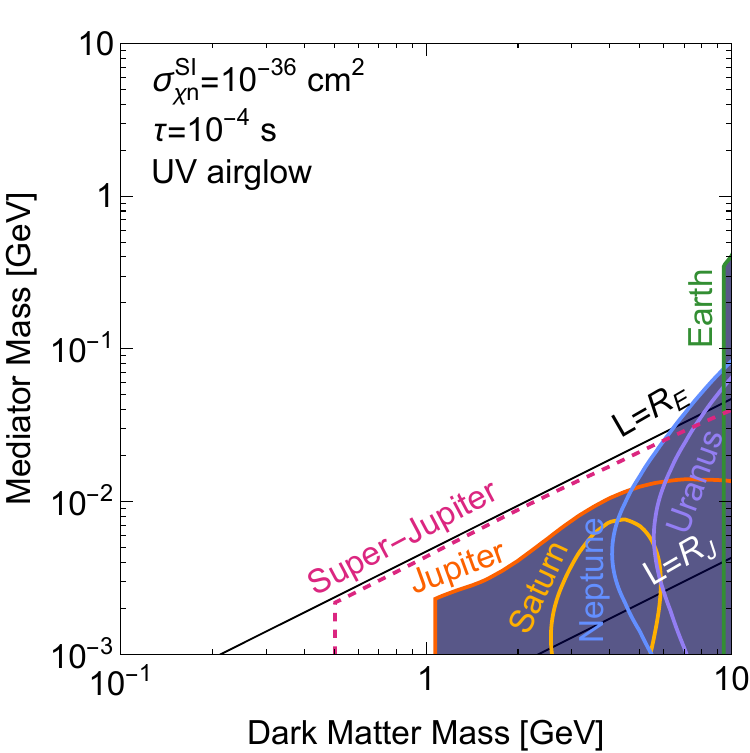}
    \includegraphics[width=0.32\textwidth]{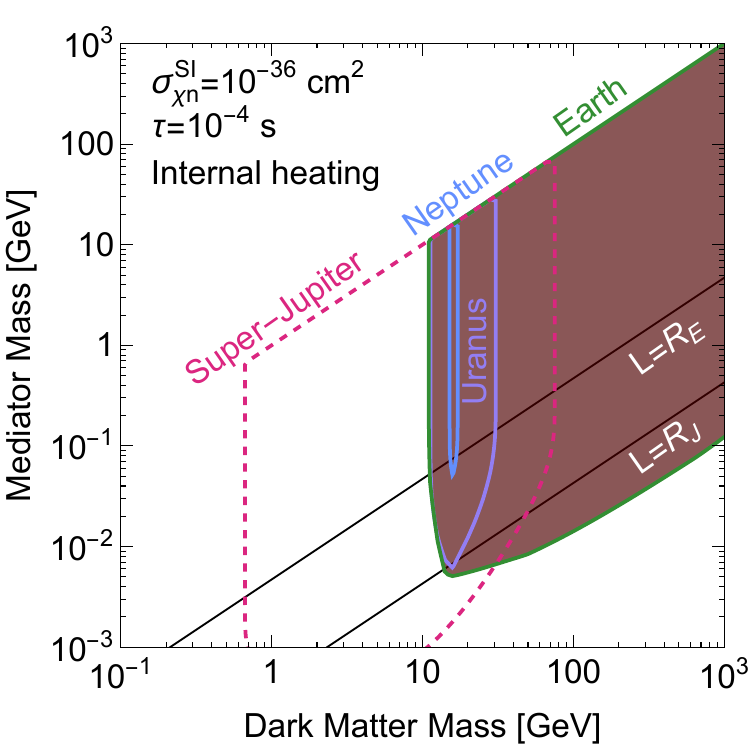}
    \caption{\label{fig:mphi_vs_mchi}Parameter space constrained by planetary UV airglow and internal heat flow for a benchmark mediator lifetime of $\tau = 10^{-4}$~s and cross sections as labeled. The lines corresponding to the choices of $L$ from the panels of Fig.~\ref{fig:fatm_light} are shown in black.}
    
\end{figure*}

Figure~\ref{fig:mphi_vs_mchi} explores the mediator and DM mass parameter space for a spin-dependent proton and spin-independent interaction of $\sigma = 10^{-36}$~cm$^2$ and mediator lifetime of $\tau = 10^{-4}$~s, for energy deposited in the atmosphere and interior. It provides the decay lengths illustrated in Fig.~\ref{fig:fatm_light}. Fig.~\ref{fig:mphi_vs_mchi_m2} of the main text presents similar results to Fig.~\ref{fig:mphi_vs_mchi}, but for a lifetime of $\tau = 10^{-2}$~s. {Comparing the results for the two lifetimes, the plot corresponds to a vertical shift in the available parameter space, as alluded to in Sec.~\ref{sec:light_xsection}. The shorter lifetime of $\tau = 10^{-4}$~s corresponds to a decay lengths becoming smaller than the typical scale of the planets, such that the parameter space where an airglow signal would be expected decreases, while the heating signal becomes dominant.}

\section{Light Mediator Energy Deposition}\label{sec:mc-detail}

\subsection{Monte-Carlo Sampling}
For each of sixty logarithmically spaced DM masses between $10^{-3}$ and $10^3$ GeV, we generate $N = 10^5$ annihilation points $\vec{x}_i$ within the planet, drawn according to either the LTE or isothermal DM density profiles. Each position is sampled with probability proportional to $n_\chi^2$, corresponding to the DM annihilation rate density. The coordinates are defined with respect to the planet's center at $\vec{0}$.

We sample the radial coordinate $r$ from the distribution $r^2 n_\chi(r)^2$ for $r \leq R_\text{atm}$, the polar angle $\Theta$ from $\sin \Theta$ over $[0, \pi]$, and the azimuthal angle $\Phi$ uniformly over $[0, 2\pi)$. At each point, we evaluate the probabilities for mediator energy deposition in both the atmosphere and the planetary interior.

\subsection{Angular Definitions}

To compute energy deposition probabilities, we define two angles, $\theta_i$ and $\varphi_i$, based on the triangle formed by the annihilation point $\vec{x}_i$, the surface point $\vec{R}_\text{atm}$, and the planetary center.

We define the path from $\vec{x}_i$ to the atmospheric surface as $\vec{s}_i = \vec{R}_\text{atm} - \vec{x}_i$, with magnitude $s_i = |\vec{s}_i|$. The relevant angles are given by
\begin{align}
\cos\theta_i &= \frac{\vec{s}_i \cdot \vec{R}_\text{atm}}{s_i R_\text{atm}}, \\
\cos\varphi_i &= \frac{\vec{x}_i \cdot \vec{R}_\text{atm}}{x_i R_\text{atm}}.
\end{align}

To determine the appropriate geometric treatment, we introduce critical angles $\theta_\text{crit}$ and $\varphi_\text{crit}$, defined by the triangle connecting the planetary center, the atmosphere-interior boundary, and a point on the atmospheric surface. These are given by:
\begin{align}
\cos\varphi_\text{crit} &= \frac{R}{R_\text{atm}}, \\
\cos\theta_\text{crit} &= \frac{\sqrt{R_\text{atm}^2 - R^2}}{R_\text{atm}}.
\end{align}

The relative values of $\cos\theta_i$ and $\cos\varphi_i$ compared to their critical values determine the classification of the annihilation point for deposition calculations.

\subsection{Geometric Regions}

We partition the space into three mutually exclusive regions, $\mathcal{A}$, $\mathcal{B}$, and $\mathcal{C}$, corresponding to different numbers of crossings of the atmosphere-interior boundary along the line from $\vec{x}_i$ to $\vec{R}_\text{atm}$.

\begin{itemize}
    \item \textbf{Region $\mathcal{A}$}: The annihilation occurs in the atmosphere, and the path to $\vec{R}_\text{atm}$ does not intersect the planetary interior. This region is defined by
    \begin{align}
        \mathcal{A} = \left\{ \cos\varphi_i \geq \cos\varphi_\text{crit} \ \land\ x_i \geq R \right\} \cup \left\{ \cos\theta_i \leq \cos\theta_\text{crit} \right\}.
    \end{align}
    
    \item \textbf{Region $\mathcal{B}$}: The annihilation occurs inside the planetary interior, $i.e.$, $x_i < R$. In this case, the path from $\vec{x}_i$ to $\vec{R}_\text{atm}$ crosses the atmosphere-interior interface exactly once.

    \item \textbf{Region $\mathcal{C}$}: The annihilation occurs in the atmosphere, but the path intersects the interior, crossing the interface twice. This region is defined as the complement of $\mathcal{A} \cup \mathcal{B}$:
    \begin{align}
        \mathcal{C} = (\mathcal{A} \cup \mathcal{B})^c.
    \end{align}
\end{itemize}

\subsection{Deposition Probabilities}

For each point, we compute the probability that the mediator decays in the atmosphere along its path to $\vec{R}_\text{atm}$. This probability depends on the region in which the point falls:

\begin{align}
\mathbb{P}^\text{atm}_i =
\begin{cases}
1 - e^{-s_i/L} & \text{if } \vec{x}_i \in \mathcal{A}, \\
e^{-(s_i - h_i^-)/L} - e^{-s_i/L} & \text{if } \vec{x}_i \in \mathcal{B}, \\
1 - e^{-(s_i - h_i^+)/L} + e^{-(s_i - h_i^-)/L} - e^{-s_i/L} & \text{if } \vec{x}_i \in \mathcal{C},
\end{cases}
\end{align}

where $L$ is the mediator decay length, and $h_i^\pm$ are the distances from $\vec{R}_\text{atm}$ to the points where the path intersects the interior boundary. These are given by:
\begin{align}
h^\pm_i = R_\text{atm} \left( \cos\theta_i \pm \sqrt{\frac{R^2}{R^2_\text{atm}} - \sin^2\theta_i} \right).
\end{align}

The corresponding interior deposition probability is:
\begin{align}
\mathbb{P}^\text{int}_i =
\begin{cases}
0 & \text{if } \vec{x}_i \in \mathcal{A}, \\
1 - e^{-(s_i - h^-_i)/L} & \text{if } \vec{x}_i \in \mathcal{B}, \\
e^{-(s_i - h^+_i)/L} - e^{-(s_i - h^-_i)/L} & \text{if } \vec{x}_i \in \mathcal{C}.
\end{cases}
\end{align}

By spherical symmetry, these probabilities are representative of the decay likelihood to any point on the atmospheric shell of radius $R_\text{atm}$.

\bibliography{main}
\bibliographystyle{apsrev4-2}

\end{document}